\documentclass{article}

\usepackage{moreverb,url}

\usepackage[colorlinks,bookmarksopen,bookmarksnumbered,citecolor=red,urlcolor=red]{hyperref}

\newcommand\BibTeX{{\rmfamily B\kern-.05em \textsc{i\kern-.025em b}\kern-.08em
T\kern-.1667em\lower.7ex\hbox{E}\kern-.125emX}}

\usepackage{amsmath}
\usepackage{amssymb}
\usepackage{a4wide}
\usepackage{float}

\usepackage{tikz}
\usetikzlibrary{calc}
\usetikzlibrary{shapes.geometric}
\usepackage{capt-of}
\usepackage{graphicx}
\usepackage{siunitx}
\usepackage{changepage}
\usepackage{subcaption}
\usepackage{textcomp}

\newcommand{\tensor}[1]{\boldsymbol{\mathrm{#1}}}

\newcommand{\opGrad}{\operatorname{Grad}}

\newcommand{\opTrace}{\operatorname{tr}}

\newcommand{\opdev}{\operatorname{dev}}
\newcommand{\body}{\mathcal{B}}

\newcommand{\tsr}[1]{\boldsymbol{\mathrm{#1}}}
\renewcommand{\vec}[1]{{\boldsymbol{#1}}}

\title{{Thermodynamically consistent modelling of viscoelastic solids under finite strain}}
\author{Mario Kunzemann, Leonhard K. Doppelbauer, Rene Preuer and Astrid Pechstein}
\begin{document}


\maketitle
\begin{abstract}
	The present article is concerned with modelling the viscoelastic behavior of Polydimethylsiloxane (PDMS) in large-strain regime. Starting from the basic principles of thermodynamics, an incremental variational formulation is derived. Within this model, the free energy density and dissipation function determine elastic and viscous properties of the solid.
	The main contribution of this paper is the estimation of the parameters in the proposed phenomenological model from measurements conducted on PDMS samples. This non-linear material model simplifies to a Prony-series representation in frequency domain in case of small deformations. The coefficients of this Prony-series are detected from dynamical temperature mechanical analysis measurements. Time-temperature superposition allows to combine measurements at different temperatures, such that a sufficiently large frequency range is available for subsequent fitting of Prony-parameters.
	A set of material parameters is thus provided. 
	The incremental variational formulation directly lends itself to finite element discretization, where an efficient and stable choice of elements is proposed for radially symmetric problems. This formulation allows to verify the proposed model against experimental data gained in ball-drop experiments. 
\end{abstract}

\section{Introduction}

Silicone-based polymers, like Polydimethylsiloxane (PDMS), are known for their high flexibility and excellent damping properties. Thermal and chemical stability, corrosion resistance and optical transparency add to the features of these materials.
Not least their easy accessiblity at low cost and simple generation of samples allow for numerous applications in science and engineering.
We cite the recent review papers by \cite{ariati2021polydimethylsiloxane} and \cite{zaman2019comprehensive}, both of which provide a great overview over the properties and applications of PDMS.

Typical applications of PDMS include, for example, microfluidics, where channels are required for filtering, separating  and mixing fluids (cf.~\cite{borok2021pdms}), the insulation of electronic components, or coating of sensitive materials to protect them from external effects (cf.~\cite{eduok2017recent}). In these applications, the damping nature of PDMS is of high advantage. An increase of the damping behavior could be realized through micro-structure design of sponges in \cite{zhu2017recent}. Treatment of surface layers was used to generate custom wrinkling patterns in specimen in \cite{knapp2021controlling}. 

The reliable realization of many of these applications requires accurate and efficient simulation techniques to predict the behavior of the system under consideration. Due to the high flexibility of PDMS, such a model must be fit to represent large deformations, and the damping properties have to be characterized properly.
\newline
%
%
Although frequently used in biomechanical or medical applications, the mechanical characterization of PDMS is not yet sufficiently accurate. Concerning the hyperelastic behavior, \cite{Nunes:2011} proposed a simple shear strain test to obtain material parameters for Mooney-Rivlin type materials. They proposed to use a variant to the classical Mooney-Rivlin material law to better fit their measurements. Later, \cite{Cardoso:2018} fitted different hyperelastic material laws, including Mooney-Rivlin, Ogden and Yeoh models. The influence of surface patterns on the elastic response was analyzed by \cite{LeeEtal:2021}. Characterizations of the damping properties are scarce, \cite{LinEtal:2009} experimentally detected Prony-parameters from punch tests and a specially developed dynamical mechanical analysis (DMA); however, the frequency range of this DMA is limited and the influence of temperature is not treated. \cite{Deguchi:2015} measured storage and loss modulus at a single frequency of $1$Hz for a range of temperatures and several PDMS polymers. In their technical note, \cite{Long:2017} treat the applicability of time-temperature-superposition to Sylgard~184.
\newline
%
We propose an approach that ensures thermodynamic consistency when dealing with large deformations in viscoelastic solids. 
The ensuing incremental formulation is based on the Clausius-Duhem inequality and is defined not only through hyperelastic energy densities, but also through a dissipation function to model viscoelastic losses.
In \cite{KunzemannEtal:2023}, a similar characterization was extended to include electromechanical coupling due to dielectric effects.
To determine the material parameters used in these functions, a specimen of Sylgard~184 is subjected to dynamical temperature mechanical analysis (DTMA) in a limited range of frequencies. Through time-temperature superposition, a sufficiently large range of frequencies can be covered. From this data, a set of material parameters for a Prony-series expansion is found through a non-negative, non-linear least squares fit, similar to \cite{kraus2017parameter}. These Prony-parameter directly correspond to the parameters of the proposed non-linear model.

In order to verify the proposed constitutive model, a finite element model of a ball-drop experiment is set up. The incremental variational formulation can directly be transferred to the context of finite element simulation. A classical stable pair of displacement-pressure elements is proposed to account for the incompressible nature of PDMS. Additional degrees of freedom for viscous strains satisfy the incompressibility condition exactly for radially symmetric problems. Time integration using the Newmark-$\beta$ scheme in variational context avoids any numerical damping.

%

%
%

The remainder of this paper is organized as follows: After the introduction, the constitutive model based on a hyperelastic potential, viscous contributions and a dissipation function is discussed. Next, the correspondence of this non-linear model to a Prony-series expansion is discussed. The estimation of Prony-parameters from DTMA measurements is conducted using time-temperature superposition as well as non-negative least squares approximation. An incremental variational formulation, which replaces D'Alembert's principle in dissipative processes, is derived and shown to be equivalent to classical viscoelastic models. This directly leads to the definition of a finite element model for radially symmetric problems in the following section. Finally, computational and experimental results are compared for a ball-drop experiment.

\section{Viscoelastic modelling} \label{sec:ViscoModel}
\subsection{Kinematics}

We consider a body $\body$ consisting of a viscoelastic material. Each material point $\mathcal{P}$ is identified by its position vector $\boldsymbol{X}$ in  reference configuration. The motion of the body is described by a time-dependent mapping $\vec \chi$, which maps each material point to its current spatial position at time $t$, $\vec{x} = \vec \chi(\vec{X},t)$. 
We use standard notations for all relevant kinematic quantities, starting with the displacement field  $\vec u = \vec x - \vec X$. 
Throughout the following, we identify derivatives of fields with respect to reference coordinates $\vec X$ by using capital operators. Derivatives with respect to spatial coordinates $\vec x$ are identified using lower case operators.

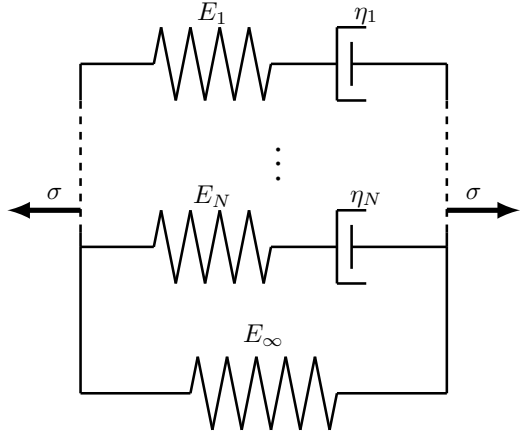
\begin{figure}[h]
		\begin{center}{} 
	\resizebox{7cm}{!}{%
		
		\begin{tikzpicture}
		
				\draw[line width=1pt] (0,0) -- (0,2){};
				\draw[line width=1pt] (0,0) -- (1.5,0){};
				\draw[line width=1pt] (1.5,0) -- (1.6,0.5){};
				\draw[line width=1pt] (1.6,0.5) -- (1.8,-0.5){};
				\draw[line width=1pt] (1.8,-0.5) -- (2,0.5){};
				\draw[line width=1pt] (2.0,0.5) -- (2.2,-0.5){};
				\draw[line width=1pt] (2.2,-0.5) -- (2.4,0.5){};
				\draw[line width=1pt] (2.4,0.5) -- (2.6,-0.5){};
				\draw[line width=1pt] (2.6,-0.5) -- (2.8,0.5){};
				\draw[line width=1pt] (2.8,0.5) -- (3.0,-0.5){};
				\draw[line width=1pt] (3.0,-0.5) -- (3.2,0.5){};
				\draw[line width=1pt] (3.2,0.5) -- (3.4,-0.5){};
				\draw[line width=1pt] (3.4,-0.5) -- (3.5,0.0){};
				\draw[line width=1pt] (3.5,0) -- (5,0){};					
				\draw[line width=1pt] (5,0) -- (5,2){};	
				\draw[line width=1pt] (0,2) -- (1,2){};

				\draw[line width=1pt] (1,2) -- (1.1,2.5){};
				\draw[line width=1pt] (1.1,2.5) -- (1.3,1.5){};
				\draw[line width=1pt] (1.3,1.5) -- (1.5,2.5){};
				\draw[line width=1pt] (1.5,2.5) -- (1.7,1.5){};
				\draw[line width=1pt] (1.7,1.5) -- (1.9,2.5){};
				\draw[line width=1pt] (1.9,2.5) -- (2.1,1.5){};
				\draw[line width=1pt] (2.1,1.5) -- (2.3,2.5){};
				\draw[line width=1pt] (2.3,2.5) -- (2.5,1.5){};
				\draw[line width=1pt] (2.5,1.5) -- (2.6,2){};
				
				\draw[line width=1pt] (2.6,2) -- (3.5,2){};
				\draw[line width=1pt] (3.5,1.5) -- (3.5,2.5){};
				\draw[line width=1pt] (3.5,1.5) -- (3.9,1.5){};
				\draw[line width=1pt] (3.5,2.5) -- (3.9,2.5){};
				\draw[line width=1pt] (3.7,1.7) -- (3.7,2.3){};
				\draw[line width=1pt] (3.7,2) -- (5,2){};
				
				\draw[line width=1pt] (0,2) -- (0,2.2){};
				\draw[line width=1pt,dashed] (0,2.2) -- (0,4){};
				\draw[line width=1pt] (0,4) -- (0,4.5){};
				\draw[line width=1pt] (5,2) -- (5,2.2){};
				\draw[line width=1pt,dashed] (5,2.2) -- (5,4){};
				\draw[line width=1pt] (5,4) -- (5,4.5){};
				
				\draw[line width=1pt] (0,4.5) -- (1,4.5){};
				\draw[line width=1pt] (1,4.5) -- (1.1,5){};
				\draw[line width=1pt] (1.1,5) -- (1.3,4){};
				\draw[line width=1pt] (1.3,4) -- (1.5,5){};
				\draw[line width=1pt] (1.5,5) -- (1.7,4){};
				\draw[line width=1pt] (1.7,4) -- (1.9,5){};
				\draw[line width=1pt] (1.9,5) -- (2.1,4){};
				\draw[line width=1pt] (2.1,4) -- (2.3,5){};
				\draw[line width=1pt] (2.3,5) -- (2.5,4){};
				\draw[line width=1pt] (2.5,4) -- (2.6,4.5){};
				
				\draw[line width=1pt] (2.6,4.5) -- (3.5,4.5){};
				\draw[line width=1pt] (3.5,4) -- (3.5,5){};
				\draw[line width=1pt] (3.5,4) -- (3.9,4){};
				\draw[line width=1pt] (3.5,5) -- (3.9,5){};
				\draw[line width=1pt] (3.7,4.2) -- (3.7,4.8){};
				\draw[line width=1pt] (3.7,4.5) -- (5,4.5){};

				\draw[-latex, line width=2pt] (5,2.5) -- (6,2.5) node[above,xshift=-18] {$\sigma$};
				\draw[latex-, line width=2pt] (-1,2.5) -- (0,2.5) node[above, xshift=-10] {$\sigma$};
				
				\node (A) at (2.5,0.4) [label=above:{${E_\infty}$}] {};
				\node (A) at (1.8,2.3) [label=above:{${E_N}$}] {};
				\node (A) at (3.9,2.3) [label=above:{${\eta_N}$}] {};
				\node (A) at (1.8,4.8) [label=above:{${E_1}$}] {};
				\node (A) at (3.9,4.8) [label=above:{${\eta_1}$}] {};
				
				\node (A) at (2.7,3) [label=above:{${\boldsymbol{\cdot}}$}] {};	
				\node (A) at (2.7,2.8) [label=above:{${\boldsymbol{\cdot}}$}] {};	
				\node (A) at (2.7,2.6) [label=above:{${\boldsymbol{\cdot}}$}] {};	

		\end{tikzpicture}
	}
	\end{center}

	\caption{Linear, one-dimensional rheological Wiechert model for viscoelastic solids.}
	\label{fig:wiechert}
\end{figure}

The deformation gradient is then defined as $\tsr F = \opGrad \vec \chi = \tsr I + \opGrad \vec u$. 
For modeling large deformation viscoelasticity, an extension of the one-dimensional rheological standard solid model \cite{Haupt2013} is proposed. 
In the standard solid model, also known as Wiechert model, a single spring element and $N$ spring-dashpot combinations in series, also known as Maxwell element, are connected in parallel, see also Figure~\ref{fig:wiechert}. 
	
For the representation of incompressible elastomers, a splitting of the deformation gradient into its volumetric and isochoric components is adopted, with a hat denoting the isochoric part $\tsr F = J^{1/3} \hat {\tsr F}$ with $J = \det \tsr F$. 
Based on the above definitions, the right Cauchy-Green tensor $\tsr C$ and isochoric component of the right Cauchy-Green tensor $\hat{\tsr{C}}$ are given by,
\begin{align}
	\tsr C &= \tsr F^\top \cdot \tsr F, \\ \hat{\tsr C} &= \hat{\tsr F}^\top \cdot \hat{\tsr F} = J^{-2/3} \tsr C.
\end{align}
For incompressible materials, its valid to say that we have $J\equiv1$ which leads to $\hat{\tsr F} \equiv \tsr F$ and therefore leads to $\hat{\tsr C} \equiv \tsr C$.

\begin{figure}[h]
\begin{center}{} 
	\resizebox{9cm}{!}{%
		\begin{tikzpicture}
			\node (A) at (0,0) [minimum width = 2.23cm, line width = 1pt,  circle, shade, draw] {};
			\node (B) at (3,-3.8) [minimum width = 2cm, minimum height = 2.5cm, line width = 1pt, rotate = 70, ellipse, shade, draw] {};
			\node (C) at (3.5,-0.7) [minimum width = 1.8cm, minimum height = 2.77cm, line width = 1pt, rotate = 60, ellipse, shade, draw] {};
			\node (D) at (6.5,0.5) [minimum width = 1.9cm, minimum height = 2.63cm, line width = 1pt, rotate = 55, ellipse, shade, draw] {};
			\node (E) at (9,2.5) [minimum width = 3.5cm, minimum height = 2cm, line width = 1pt, rotate = -33, ellipse, shade, draw] {};
			\node at (3.1,-2.2) [rotate = 90] {\ldots};
			\draw[-latex, line width = 1.5pt] (A) to [bend right = 30] node[below left ] {$\tsr{F}_{v,N}$} (B);
			\draw[-latex, line width = 1.5pt] (A) to [bend right = 20] node[above] {$\tsr{F}_{v,1}$} (C);
			\draw[-latex, line width = 1.5pt] (A) to [bend left = 20] node[above] {$\hat{\tsr{F}}$} (D);
			\draw[-latex, line width = 1.5pt] (C) to [bend right = 30]  node[above left = -1mm] {$\tsr{F}_{e,1}$} (D);
			\draw[-latex, line width = 1.5pt] (D) to [bend right = 20] node[below right] {$J^{1/3} \tsr{I}$} (E);
			\draw[-latex, line width = 1.5pt] (A) to [bend left = 20] node[above] {$\tsr{F}$} (E);
			\draw[-latex, line width = 1.5pt] (B) to [bend right = 20] node[below right] {$\tsr{F}_{e,N}$} (D);
		\end{tikzpicture}
		}
	\end{center}
		\caption{Representation of the multiplicative decomposition.}
		\label{fig:multDecomp}
\end{figure}
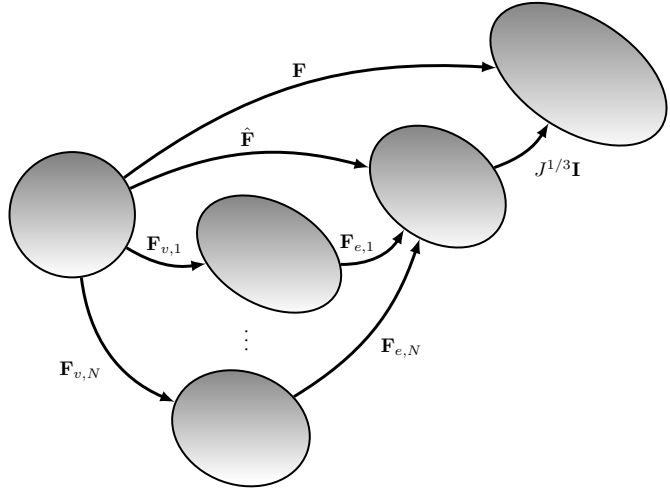

At large deformations, as an extension of the linear model, $N$ independent stress-free, volume-preserving \emph{intermediate} configurations are postulated, to which the respective elastic material laws can be applied. For each of these configurations, the isochoric part of the deformation gradient decomposes multiplicatively into an elastic part $\hat{\tsr F}_{e,i}$ and a viscoelastic part $\tsr F_{v,i}$, see Figure~\ref{fig:multDecomp}, or see e.g., \cite{ask2012electrostriction},
\begin{equation}
	\hat{\tsr F} = \hat {\tsr F}_{e,i} \cdot \tsr F_{v,i}, \text{ for }
	i \in \{1, \dots N\}.
\end{equation}
Above, the viscoelastic parts $\tsr F_{v,i}$ of the deformation resemble the transformation to the stress-free $i^{th}$ intermediate configuration, where $\tsr F_{v,i}$ is not necessarily compatible. However, it is assumed that the viscoelastic contributions are modeled in an isochoric way, such that $J_{v,i} = \det \tsr F_{v,i} \equiv 1$ for all $i = 1 \dots N$.
Based on the $i^{th}$ elastic isochoric part of the deformation gradient $\hat{\tsr F}_{e,i} = \hat {\tsr F}\cdot \tsr F_{v,i}^{-1}$, the corresponding elastic and viscoelastic Cauchy-Green tensors can be defined via,
\begin{align}
	\hat {\tsr C}_{e,i} &= \hat {\tsr F}_{e,i}^\top \cdot \hat{\tsr F}_{e,i} =
	{\tsr F}^{-\top}_{v,i} \cdot \hat{\tsr C} \cdot {\tsr F}_{v,i}^{-1}, \\
	\tsr C_{v,i} &= \tsr F_{v,i}^\top \cdot \tsr F_{v,i}.
\end{align}

\subsection{Constitutive equations}

In the following, a thermodynamically consistent model for a viscoelastic material shall be developed.
To this end, the free energy density $\Psi$ shall be defined for  independent Cauchy-Green tensor $\tsr C$ or deformation gradient $\tsr F$.
The viscoelastic Cauchy-Green tensors $\tsr C_{v,i}$ will serve as internal variables, and pressure $p$ will act as a Lagrangian multiplier for the incompressibility constraint.

The free energy is assumed to be additively composed of several independent potentials, compare \cite{ask2012electrostriction}. We consider a decomposition into a hyperelastic potential $\Psi_\infty$ and a viscoelastic energy densities $\Psi_{v,i}$, which shall be defined in the following,
\begin{equation}
	\Psi(\tsr C, \tsr C_{v,i}, p) =
	\Psi_\infty({\tsr C}, p) + \sum_{i=1}^N \Psi_{v,i}(\hat{\tsr C}, \tsr C_{v,i}) .
	\label{eq:psi_additive}
\end{equation}
For the formulation of the constitutive equations we use the framework of the Clausius-Duhem inequality.
This inequality states essentially that the dissipation $\mathcal{D}$ is non-negative (cf. \cite{Doghri2013}), which can be written as,  
	\begin{align}
		\label{eq:dissipation}
		\mathcal D &= \frac{1}{2} \tsr T : \dot{\tsr C}  - \dot \Psi  \geq 0.
	\end{align}
	Above, $\tsr{T}$  is the second Piola-Kirchhoff stress. Computing the rate $\dot \Psi$ explicitly, equation \eqref{eq:dissipation} becomes,
	\begin{equation}
		\mathcal D =-\left( \frac{\partial \Psi}{\partial \tsr C} - \frac{1}{2} \tsr T \right): \dot{\tsr C} 
		- \sum_{i=1}^N \frac{\partial \Psi}{\partial \tsr C_{v,i}} : \dot{\tsr C}_{v,i} 
		- \frac{\partial \Psi}{\partial p} \dot p  \geq 0.
		\label{eq:dissipation2}
	\end{equation}
	Equation \eqref{eq:dissipation2} implies the classical constitutive equations for the second Piola-Kirchhoff stress $\tsr T$  as,
	\begin{align}
		\tsr T = 2 \frac{\partial \Psi}{\partial \tsr C}. 
	\end{align}
	Moreover, pressure $p$ acting as a Lagrangian multiplier ensures the condition of incompressibility as $\partial \Psi /\partial p = 0$.
	The reduced dissipation inequality states that the negative inner product of \emph{generalized driving forces} $\;\partial \Psi/\partial \tsr C_{v,i}$ and viscous strain rates $\dot{\tsr C}_{v,i}$ is non-negative,
	\begin{align}
		\mathcal D &=
		- \sum_{i=1}^N \frac{\partial \Psi}{\partial \tsr C_{v,i}} : \dot{\tsr C}_{v,i} 
		\geq 0.
	\end{align}

	\paragraph{Free energy}
	Let us collect specific representations for the energy densities constituting the augmented free energy $\Psi$.
	The elastic part of the free energy $\Psi_\infty$ will be governed by an incompressible Neo-Hookean hyperelastic potential. We use $\mu_\infty = G_\infty$ for the long-term shear modulus,
	\begin{equation}
		\label{eq:psineohookep}
		{\Psi}_\infty({\tsr{C}},p) = \frac{\mu_\infty}{2}\left(\opTrace{\mathbf{C}} - 3 \right) - \mu_\infty \log J + p \log J .
	\end{equation}
	Pressure $p$ serves as a Lagrangian multiplier to ensure the incompressibility constraint, $$\frac{\partial \Psi_\infty}{\partial p} = \log J = 0,$$ such that $J \equiv 1$.
	
	Concerning the $i^{th}$ multiplicative decomposition, a similar incompressible Neo-Hookean hyperelastic potential models the elastic energy for $\hat{\tsr C}_{e,i}$ with shear modulus $\mu_{v,i}$.
	As $\opTrace \hat{\tsr C}_{e,i} = \hat{\tsr C}:{\tsr C}_{v,i}^{-1}$, we use the representation,
	\begin{equation}
		\label{eq:psiVisco}
		\Psi_{v,i}({\hat{\tsr{C}},\mathbf{C}_{{v},i}})  
		=\frac{\mu_{v,i}}{2}\left(\hat{\tsr C}:\tsr C_{v,i}^{-1} - 3 \right).
	\end{equation}

	\paragraph{Dissipation}
	Viscoelastic evolution is defined implicitly through a \emph{dissipation function} $\Phi$ such that,
	\begin{equation}
		\mathcal D = \sum_{i=1}^N\frac{\partial \Phi}{\partial \dot{\tsr C}_{v,i}} : \dot{\tsr C}_{v,i}.
	\end{equation}
	As for the free energy densities, we also assume the dissipation function to be composed additively from contributions due to the $N$ internal variables $\tsr C_{v,i}$. Each of the terms is governed by a viscosity parameter $\eta_i$. To be specific, we use,
	\begin{align}
		\Phi = \sum_{i=1}^N \Phi_i(\tsr{C}_{v,i},\dot{\tsr{C}}_{v,i}),
	\end{align}
with,
	\begin{align}
		\Phi_i(\tsr{C}_{v,i},\dot{\tsr{C}}_{v,i}) = \frac{1}{12} \sum_{i=1}^N \eta_i ({\tsr C}_{v,i}^{-1}  \cdot \dot{\tsr C}_{v,i} ):({\tsr C}_{v,i}^{-1}  \cdot \dot{\tsr C}_{v,i} ).
		\label{eq:def_phi}
	\end{align}


To analyze the implications of free energy $\Psi$ and dissipation function $\Phi$ on the evolution of $\tsr{C}_{v,i}$, an incremental variational formulation based on these quantities is posed later. There, we see that classical viscoelastic material models are reproduced for the above choices. Additionally, the incremental variational formulation can directly be transferred to a finite element discretization, which we do for the special case of rotationally symmetric problems.

\section{Temperature-time shifting and parameter fitting of measured data}
\label{sec:TTS}
\subsection{Linear viscoelasticity }
	
The viscoelastic parameters such as the long-term modulus $\mu_\infty$ and the viscous properties $\mu_i$ and $\eta_i$  of our non-linear constitutive model
are characterized from small-strain measurements, such that an identification using the linear Wiechert model is valid. 
Within this model, each spring is characterized through its elastic modulus $E_i$. The long-term modulus of the system equals $E_\infty$, i.e. the modulus of the single spring. Its instantaneous modulus $E_0$ can be found as, 
\begin{align} 
	E_0 = \sum_{i=1}^{N} E_i + E_\infty,
	\label{eq:E0}
\end{align}
for a thorough discussion we refer to \cite{Haupt2013}.
For incompressible materials with a Poisson ratio of $\nu=0.5$, shear moduli $\mu_i$  and Young's moduli $E_i$ are connected via,
\begin{align}
	\label{eq:mui}
	\mu_i &= \frac{E_i}{1+2\nu} = \frac{E_i}{3}, & 
	i \in \{1, \dots N, \infty\}.
\end{align} 
Additionally, $\tau_i$ defines the relaxation time of the $i^\text{th}$ Maxwell element by,
\begin{align}
	\eta_i = \tau_i E_i.
	\label{eq:etai} 
\end{align} 

At time $t > 0$, the relaxation modulus is found as, see  \cite{Brinson2015},
\begin{align}
	E(t) = \sum_{i=1}^{N}E_i e^{-t/\tau_i} + E_\infty.
	\label{eq: relaxE}
\end{align}
Representation \eqref{eq: relaxE} is often referred to as Prony-series. After Fourier transformation to frequency domain, one obtains the more common series expansion,
\begin{align}
	E^*(\omega) = \underbrace{E_\infty + \sum_{i=1}^{N} \frac{E_i \tau_i^2 \omega^2}{1+\tau_i^2 \omega^2}}_{\substack{E'(\omega)}} + j \underbrace{\sum_{i=1}^{N} \frac{E_i \tau_i \omega}{1+\tau_i^2 \omega^2}}_{\substack{E''(\omega)}},
	\label{eq:pronySeries}
\end{align}
where $E^*(\omega)$ is termed the complex modulus of the material. 
In this context, the real part $\Re(E^*(\omega))$ is called storage modulus $E'(\omega)$ and the imaginary part $\Im(E^*(\omega))$ is known as the loss modulus $E''(\omega)$, see \cite{Tschoegl2012}. 

%
%
%
Furthermore we define $e_i = E_i/E_0$ with the instantaneous modulus $E_0$ as defined in \eqref{eq:E0}. Following our model, the material is now fully characterized when estimating the $2N$ parameters $e_i$ and $\tau_i$ as well as the long-term modulus $E_\infty$. We often refer to $N$ as the \emph{order} of the Prony-series.

%
%



%
\subsection{Dynamical temperature mechanical analysis}
\begin{figure}
	\centering
	
	\begin{subfigure}{0.49\textwidth}
		\centering
		\includegraphics[width=\textwidth]{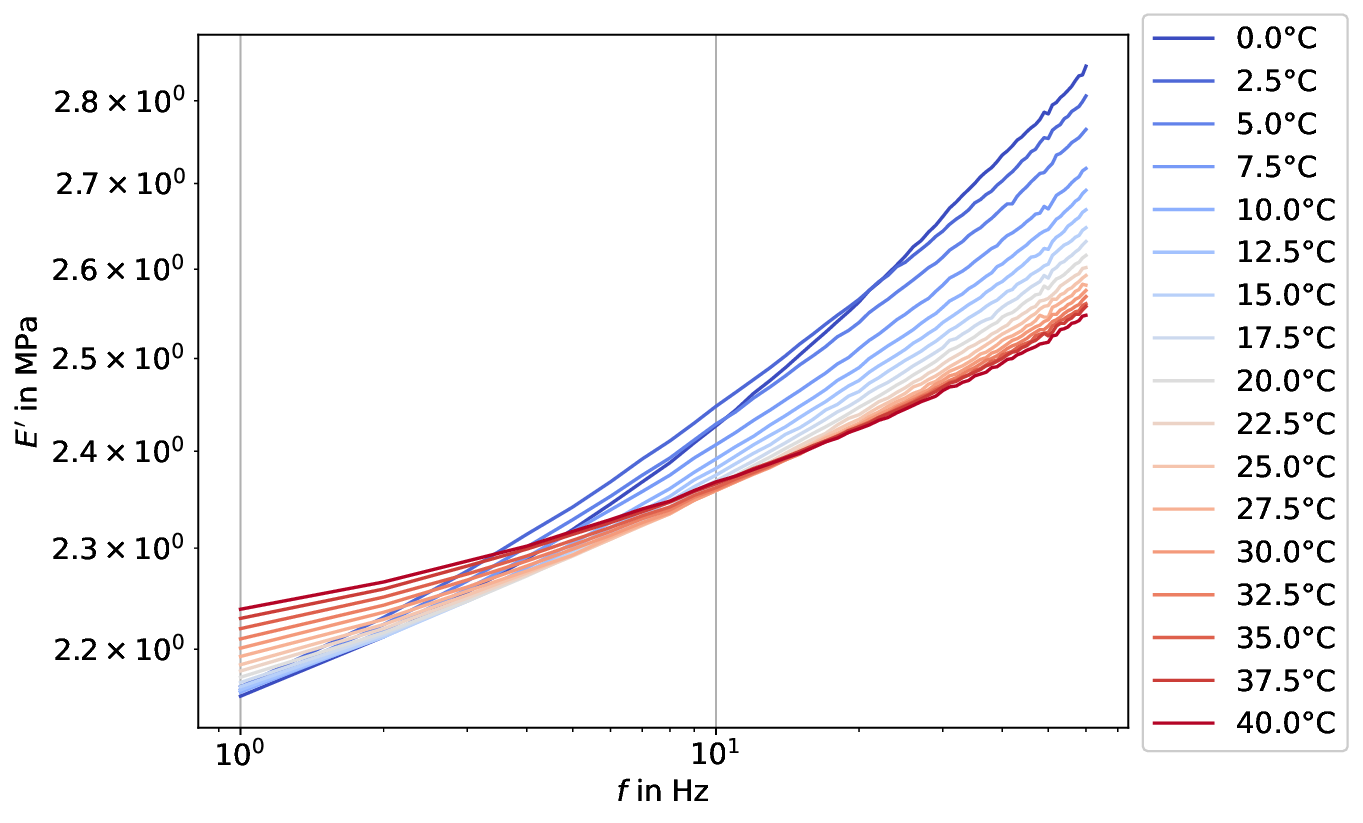}
		\caption{Measured storage modulus $E'$.}
		\label{fig:leodata-estore}
	\end{subfigure}
	\begin{subfigure}{0.49\textwidth}
		\centering
		\includegraphics[width=\textwidth]{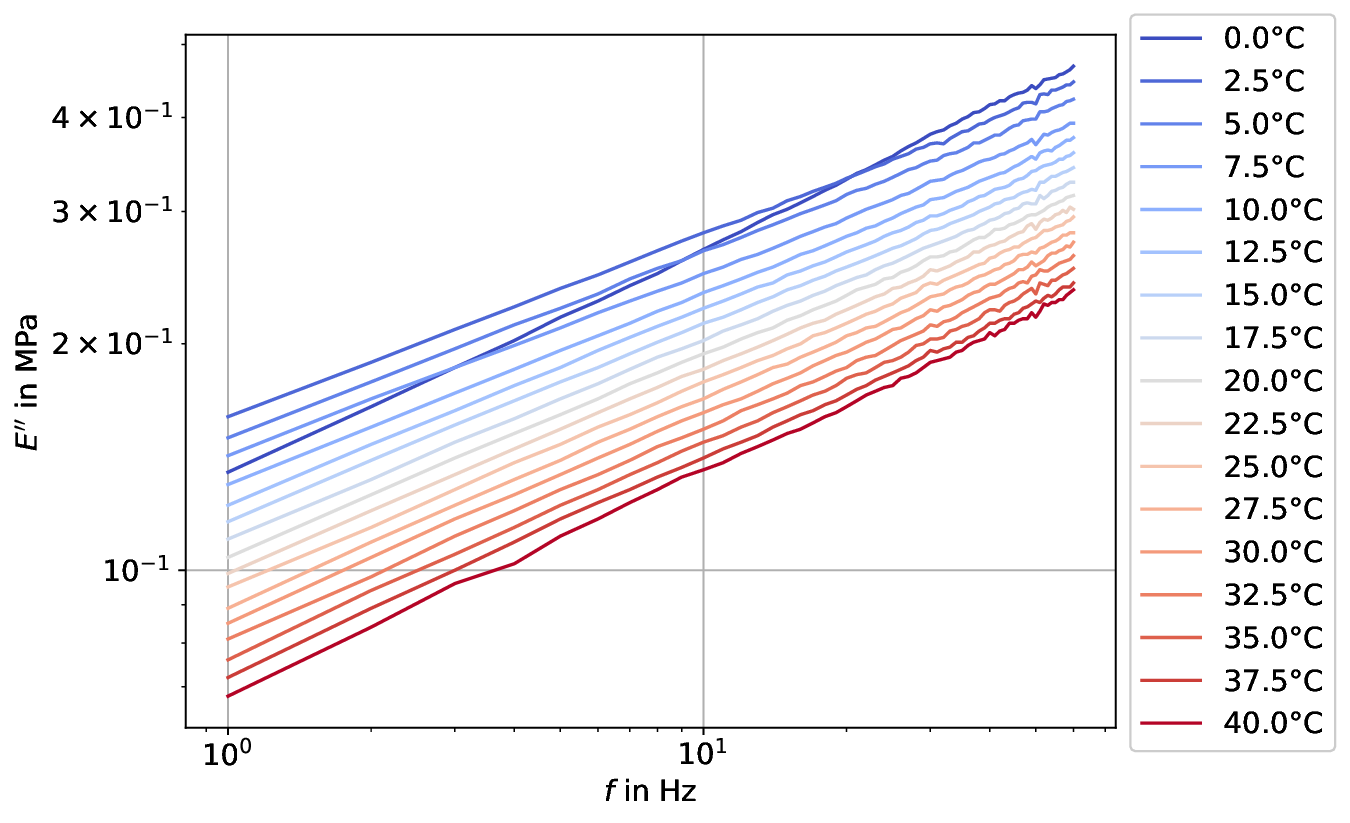}
		\caption{Measured loss modulus $E''$.}
		\label{fig:leodataeloss}
	\end{subfigure}
	\caption{DTMA measured data}

	\label{fig:dtmaMeas}
\end{figure}

To determine these Prony-parameters from experiments, dynamical temperature mechanical analysis (DTMA) measurements have been performed. 
These DTMA measurements provide storage and loss modulus over a given frequency range at different temperatures.
The measurements were carried out on a Netzsch DMA GABO \textit{Eplexor}\textregistered 500N~HT. The dimensions of the Sylgard~184 sample were $\SI{10}{\milli\meter} \times \SI{40}{\milli\meter} \times \SI{3,2}{\milli\meter}$ ($W\times H\times D$) and a clamping length of $\SI{20}{\milli\meter}$ was chosen. 
The static strain was $\SI{3}{\percent}$, the dynamic strain $\SI{2}{\percent}$ with a control range of $\SI{0.0045}{\percent}$ and the pre-stress was $\SI{0.05}{\mega\pascal}$. The pre-cooling time in the temperature chamber was at least $2$ minutes. Measurements were taken in the range $0$ to $\SI{40}{\degreeCelsius}$ in $\SI{2.5}{\degreeCelsius}$ steps between $0$ to $\SI{100}{\hertz}$. Each measurement contains $12$ cycles, from which $10$ are used to compute the average. 
Due to resonance peaks for all curves at around $\SI{72}{\hertz}$, we did not use the entire measurement curve, but only from $\SI{0}{\hertz}$ to $\SI{60}{\hertz}$, as shown in Figure~\ref{fig:leodata-estore} and Figure~\ref{fig:leodataeloss}.

This range is too small to describe the whole range of frequencies excited in applications. The time-temperature superposition principle (TTSP) allows to generate a so-called master curve that can cover a much larger frequency spectrum from measurements at different temperatures. Conversely, predictions can also be made about the long-term behaviour, see also \cite{Brinson2015}, \cite{Ferry1980}.

\subsection{Time-temperature superposition}
Concerning time-temperature superposition, Sylgard~184 is treated as a thermorheologically complex material, cf.~\cite{Klompen}.
In the modeled area, the cross-linked material is well above the glass transition temperature $T_g$ and the melting temperature $T_m$, cf.~\cite{dewimille2005synthesis}. This results in a melt-like mobility, which can be described by means of thermorheologically complex modeling.
The selected approach combines the classic TTSP for solid polymers (complex Young's modulus) and the TTSP for polymer melts (complex viscosity).
Thus, horizontal frequency shifts as well as vertical shifts in the modulus are applied.
In \cite{Tschoegl2002TheEO}, the following general form of time-temperature shifts was identified as,
\begin{align}
	E_{T}^{*}(\omega) = b_T  E_{T_0}^{*}(a_T\omega).
	\label{eq:TTSP}
\end{align} 
Shifting parameters $a_T$ and $b_T$ allow to transfer the material's modulus from reference temperature $T_0$ to any other temperature $T$, at least within a certain temperature range.
%
A simple approach for shifting the modulus is given by,
\begin{align}
	b_T = \frac{T_0 \rho_0}{T \rho},
	\label{eq:bT}
\end{align}
where $\rho $ and $\rho_0$ are the mass density of the polymer at $T$ and $T_0$. 
Since we assume no relevant density differences within the sampled temperature range of $\SI{0}{\degreeCelsius}$ to $\SI{40}{\degreeCelsius}$, this density ratio is neglected in the following.  

For the horizontal frequency shift a large variety of models can be found. Probably the one most commonly used is the Williams-Landel-Ferry model (WLF). 
However, a main criterion for this is free-volume availability, which is despite the amorphous morphology at ambient temperature limited by the intermolecular bonds created in cross linking. As shown in \cite{LOMELLINI19924983}, the WLF model can be used very well for a temperature range of $T_g < T < T_g +\SI{100}{\degreeCelsius}$.  

\cite{CLARSON1985930} and also \cite{LI2006580} found that the glass transition temperature of Sylgard~184 is in the range of $ \SI{-150}{\degreeCelsius}$ $ < T_g < \SI{-120}{\degreeCelsius}$ and the melting temperature $T_m$ lies between $\SI{-50}{\degreeCelsius}$ and $\SI{-70}{\degreeCelsius}$. This implies that the WLF model is not applicable to Sylgard~184 at ambient temperatures.
\cite{LOMELLINI19924983} states that in this relatively high temperature range of $T>T_g+\SI{100}{\degreeCelsius}$, the so-called Arrhenius equation is applicable,
\begin{align}
	a_T = e^{\alpha\left(\frac{1}{T}-\frac{1}{T_0}\right)}.
	\label{eq:aT}
\end{align}
The parameter $\alpha = E_a/\mathcal{R}$ is independent of temperature, but can be computed from the activation energy $E_0$ and the universal gas constant $\mathcal{R}$. We use the DTMA measurement data to compute $\alpha$, and thereby the activation energy $E_a$, using a least-squares method.

\begin{figure}
	\centering
	\begin{subfigure}{0.49\textwidth}
		\centering
		\includegraphics[width=\textwidth]{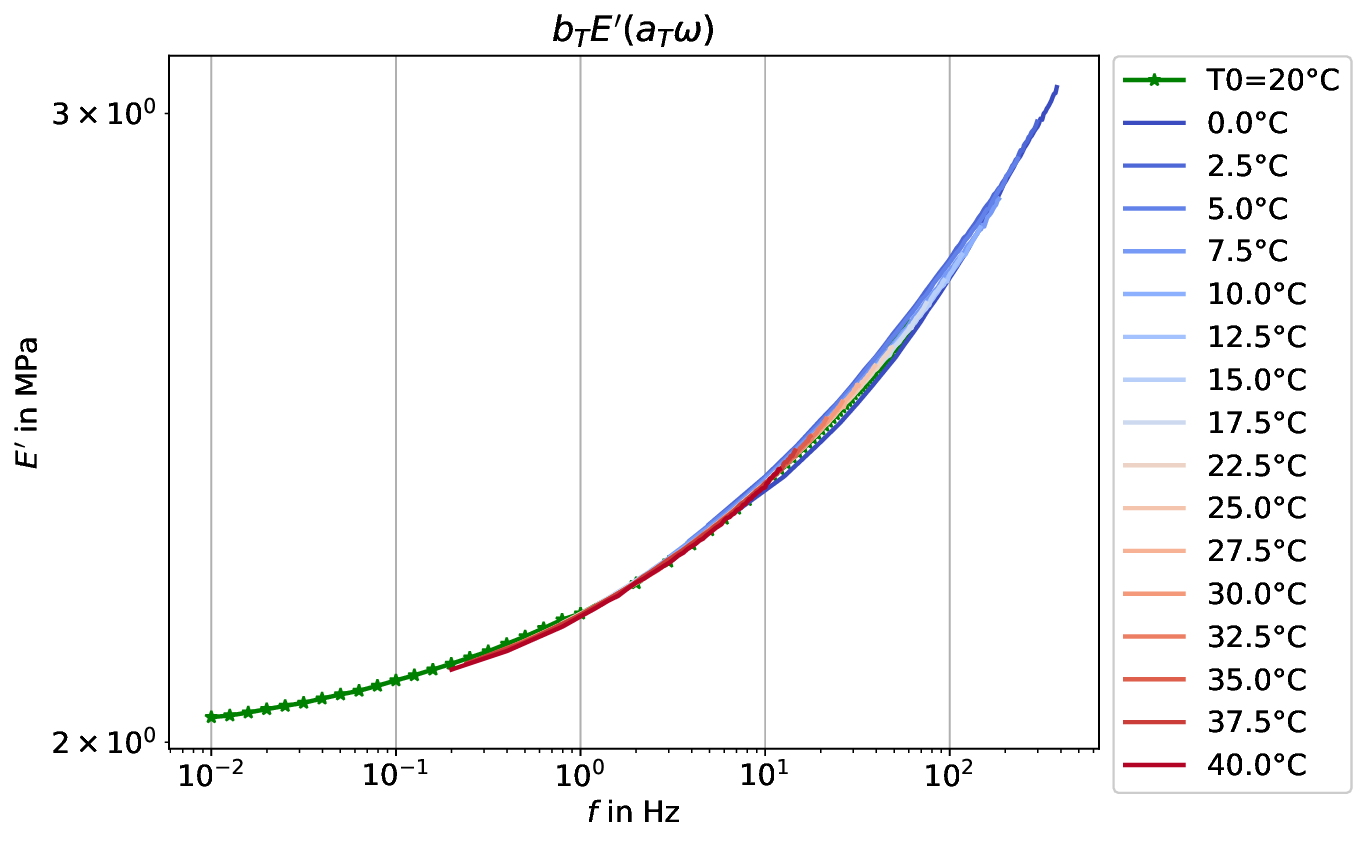}
		\caption{Storage modulus $E_{T_0}'$ at $T_0 = \SI{20}{\degreeCelsius}$.}
		\label{fig:shiftestore}
	\end{subfigure}
	\begin{subfigure}{0.49\textwidth}
		\centering
		\includegraphics[width=\textwidth]{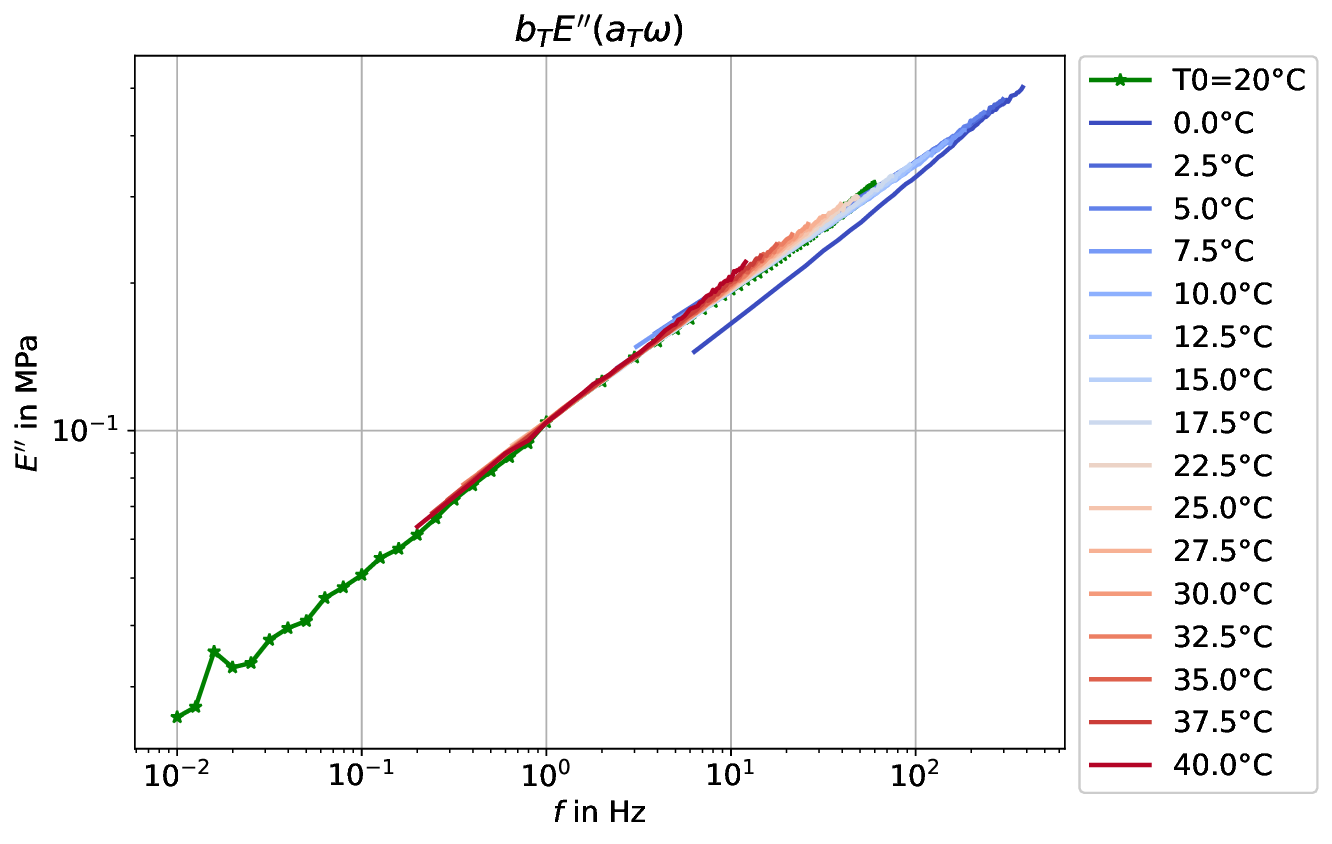}
			\caption{Loss modulus $E_{T_0}''$ at $T_0 = \SI{20}{\degreeCelsius}$.}
		\label{fig:shifteloss}
	\end{subfigure}
	\caption{Master curves for storage and loss modulus at $T_0 = \SI{20}{\degreeCelsius}$ gained through time-temperature superposition.}

	\label{fig:shiftData}
\end{figure}

In order to estimate $\alpha$, at reference temperature of $T_0 = \SI{20}{\degreeCelsius}$ measurements within an extended frequency range were conducted. While it is of course not possible to extend this range towards higher frequencies, measurements for lower frequencies approaching zero could easily be conducted; the results are displayed in Figure~\ref{fig:shiftData}.
Then, data sets for temperatures from $\SI{20}{\degreeCelsius}$ to $\SI{40}{\degreeCelsius}$ were used to determine the unknown shift parameter $\alpha$, as higher temperatures correspond to lower frequencies.
A least squares method is used to find the optimal parameter $\alpha$  in order to minimise the squared vertical distance of the respective data curves when shifted by $b_T$ and $a_T$ as in \eqref{eq:bT} and \eqref{eq:aT}, respectively. 
As the shifted frequencies for the data points do no coincide in general,
a linear interpolation of the reference data $E^*_{T_0}$ in between data points is used instead. For $\omega_i, i = 1 \dots n_\omega$ the frequencies at which DTMA data is available, the least-squares problem finally reads,
\begin{align}
	\sum_{T > T_0} \sum_{i=1}^{n_\omega} \left| b_T E^*_{T_0}(a_T \omega_i) - E^*_T(\omega_i) \right|^2 \to \min_{\alpha > 0}.
\end{align}
For the available DTMA data, this procedure results in $\alpha=\SI{7389.124}{\kelvin}$. 
Then, both the storage and loss moduli from all DTMA data temperatures can be shifted according to \eqref{eq:TTSP} to get a master curve at temperature $T_0 = \SI{20}{\degreeCelsius}$ as shown in Figure~\ref{fig:shiftData}. These shifted data sets are used in the subsequent detection of Prony-parameters; the additional data in the extended frequency range for temperature $T_0$ are not included in this process anymore.

\subsection{Prony-parameters}
Once data in an extended frequency range has been generated using the TTSP, Prony-parameters $E_\infty$ as well as $e_i$ and $\tau_i$ can be detected. To this end, a non-linear least-squares optimization problem for the $2N+1$ unknown Prony-parameters needs to be solved.
For the presented data, the non-negative least squares based routine \texttt{curve\_fit} provided within python's \texttt{scipy} package could be applied successfully.

This function works with complex data sets, and thereby ensures that the parameters determined are accurate for both the storage and loss modulus.
Moreover, range conditions have to be met for the different parameters. A minimal set of these conditions is to seek for
\begin{itemize}
	\item a positive long-term modulus $E_\infty > 0$,
	\item positive relaxation times $\tau_i>0$,
	\item relative viscoelastic moduli $0 < e_i < 1$,
	\item satisfying additionally $\sum_{i=1}^{N} e_i \leq 1 $.
\end{itemize}
Simple range conditions like the former three items can be included in the parameters passed to the least-squares fitting routine. 
The latter condition has to be verified manually, once the parameters are computed. As one often observes in non-linear problems, choosing a feasible starting value is of utmost importance. 
We observed that setting all values to one worked well when seeking the long-term modulus equivalent to $\SI{1}{\mega \pascal}$.

\begin{figure}[h]
	\centering
	\includegraphics[width=0.69\linewidth]{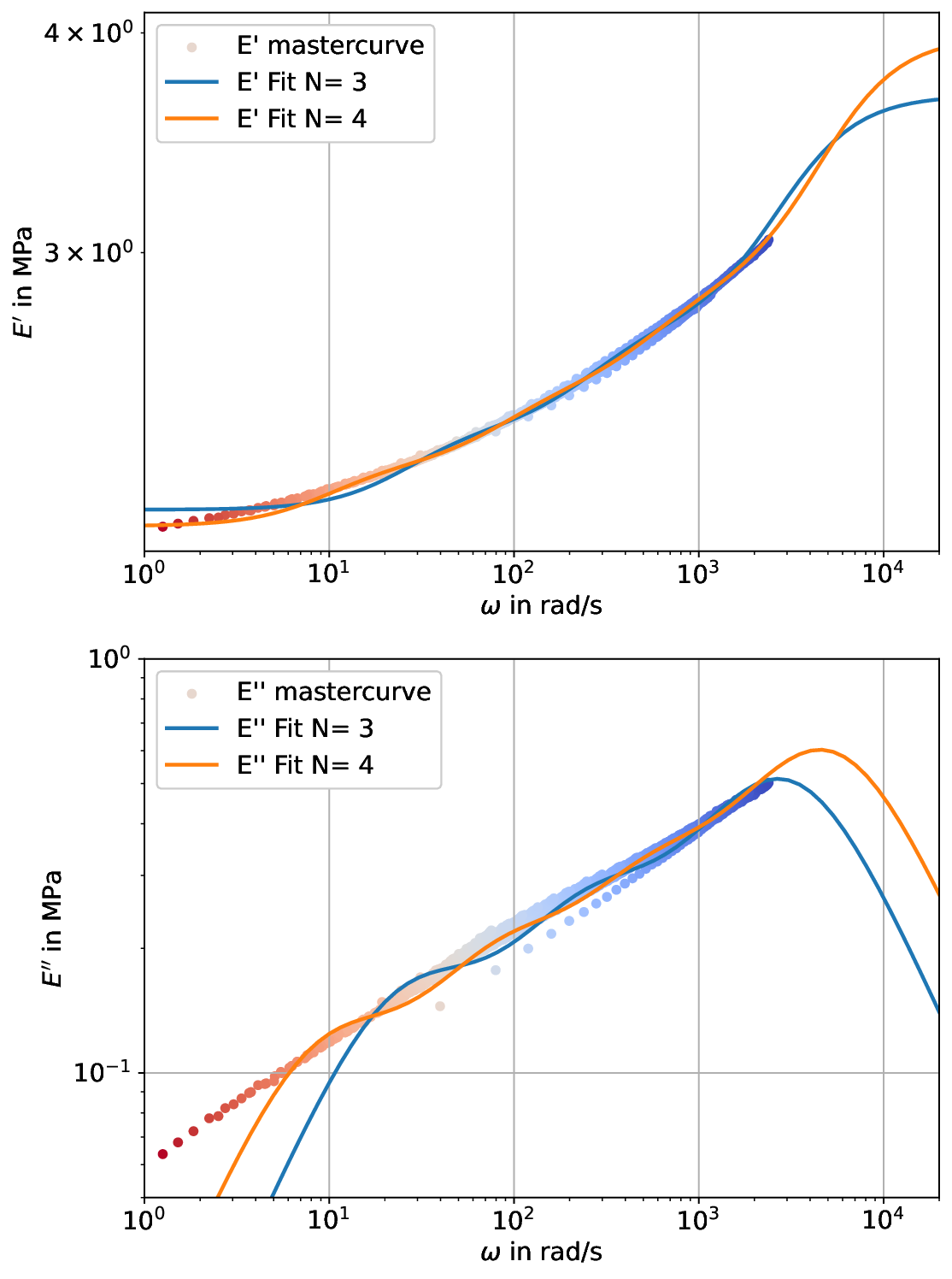}
	\caption{Fitted Prony-series, comparison $N=3$ and $N=4$.}
	\label{fig:pronyfit}
\end{figure}

Finally, the order $N$ of the Wiechert model has to be chosen in accordance with the available data set.
As a rule of thumb, good approximation is obtained when using one Maxwell element per measured frequency decade.  
Since after TTSP, measurements are available for frequencies from $\SI{0.1}{\hertz}$ up to almost $\SI{1000}{\hertz}$ (see Figure~\ref{fig:shiftData}), fitting Prony-parameters for models of third and fourth order seems appropriate.
The respective results are depicted in Figure~\ref{fig:pronyfit}. 
Table~\ref{tab:paraO3} and Table~\ref{tab:paraO4} show the corresponding Prony-parameters $E_\infty$, $E_i = e_i E_0$ and $\tau_i$ as well as the shear moduli $\mu_i$ and viscosities $\eta_i$ computed from these values via \eqref{eq:mui} and \eqref{eq:etai}.

\begin{table}[h]
	\footnotesize
	\centering
	\caption{Prony-parameters for $N = 3$ and $T_0 = \SI{20}{\degreeCelsius}$. }
	\label{tab:paraO3}
	\begin{tabular}{ccccc}
		\hline
		$i$ & $E_i$ in $\si{\mega \pascal}$ &$\tau_i$ in $\si{\second}$ & $\mu_i$ in $\si{\mega \pascal}$   & $\eta_i$ in $\si{\mega \pascal \second}$ \\
		\hline
		$\infty$ & $2.11904$ & - & $0.70635$ & - \\
		\hline
		1 & $0.93789$ & $0.00037$ & $0.31263$ & $0.00035$ \\
		\hline
		2 & $0.34398$ & $0.00424$ & $0.11466$ & $0.00146$ \\
		\hline
		3 & $0.24635$ & $0.04684$ & $0.08212$ & $0.01154$ \\
		\hline
	\end{tabular}
\end{table}

\begin{table}[h]
	\footnotesize
	\centering
	\caption{Prony-parameters for $N = 4$ and $T_0 = \SI{20}{\degreeCelsius}$.}
	\label{tab:paraO4}
	\begin{tabular}{ccccc}
		\hline
		$i$ & $E_i$ in $\si{\mega \pascal}$ &$\tau_i$ in $\si{\second}$ & $\mu_i$ in $\si{\mega \pascal}$   & $\eta_i$ in $\si{\mega \pascal \second}$ \\
		\hline
		$\infty$ & $2.07639$ & - & $0.69213$ & - \\
		\hline
		1 & $1.06895$ & $0.00023$ & $0.35632$ & $0.00024$ \\
		\hline
		2 & $0.34258$ & $0.00198$ & $0.11419$ & $0.00068$ \\
		\hline
		3 & $0.24093$ & $0.01323$ & $0.08031$ & $0.00319$ \\
		\hline
		4 & $0.17836$ & $0.1386$ & $0.05945$ & $0.02472$ \\
		\hline
	\end{tabular}
\end{table}

In both Table~\ref{tab:paraO3} and Table~\ref{tab:paraO4}, we note that the relaxation times $\tau_i$ are distributed more or less evenly across the measured decades, as is to be expected due to \cite{Tschoegl2012}.

\section{Incremental minimization principle} \label{sec:Increment}

%
Having determined the relevant constitutive parameters, we now seek a finite element formulation for the problem at hand. As a starting point, a variational formulation for viscoelastic solids is derived from an incremental minimization problem. A similar procedure has been presented for modeling various dissipative processes by the group around Miehe, we cite non-exhaustively \cite{Miehe:2002,MieheEtal:2002,miehe2011variational}.
%
For reasons of simplicity of presentation, we restrict ourselves to the case of a conservative volume force density $\vec f$. With $\rho_0$ the mass density, the kinetic energy density $T$ and local power of external forces $\mathcal{P}_{ext}$ read,
\begin{align}
		T &= \frac{1}{2} \rho_0 \dot{\vec{u}}\cdot \dot{\vec{u}}, &
		\mathcal{P}_{ext} &= \vec f \cdot \dot {\vec u}.
\end{align}

Consider a finite time step $[t_0, t_0+\Delta t]$ with known initial conditions at time $t_0$. 
We introduce the \emph{incremental potential} for this time step as,
\begin{align}
\begin{split}
		\Pi_t^{t+\Delta t} &= \int_{t_0}^{t_0+\Delta t}( \dot T + \dot \Psi + \Phi - \mathcal P_{ext})\, dt\\
	&= \underbrace{T_{t_0+\Delta t} - T_{t_0}}_{\text{kinetic}} +
	\underbrace{\Psi_{t_0+\Delta t} - \Psi_{t_0}}_{\text{stored}} + 
	\underbrace{\int_{t_0}^{t_0+\Delta t} \Phi dt}_{\text{dissipated}}  - 
	\underbrace{\int_{t_0}^{t_0+\Delta t} \mathcal P_{ext}\, dt}_{\text{external}}.
\end{split}
\end{align}
The incremental potential $\Pi_{t_0}^{t_0+\Delta t}$ is then minimized with respect to all admissible displacements and viscoelastic internal parameters, and maximized with respect to hydrostatic pressure. 
For time step size $\Delta t$ approaching zero, we arrive at,
\begin{align}
	\label{eq:minPro1}
	\dot T + \dot \Psi +  \Phi - \mathcal P_{ext} \to \min_{\dot {\vec{u}}} \min_{\dot{\tsr C}_{v,i}}  \max_{\dot p}.
\end{align}
We can now rewrite equation \eqref{eq:minPro1} for the specific choice of free energy $\Psi$ and get, 
\begin{equation}
	\begin{split}
		\rho  \ddot{\boldsymbol{u}} \cdot \dot{\boldsymbol{u}}
		+ \frac{\partial \Psi}{\partial \tsr{C}}\colon \dot{\tsr{C}}
		+ \sum_{i=1}^{N}\frac{\partial \Psi}{\partial \tsr{C}_{v,i}}\colon \dot{\tsr{C}}_{v,i}
		+ \frac{\partial \Psi}{\partial {p}} \dot{p} 
		+ \Phi - \vec{f} \cdot \vec{\dot{u}}
		\to
		\min_{\dot {\vec{u}}} \min_{\dot{\tsr C}_{v,i}}  \max_{\dot p}.
	\end{split}
\end{equation}
Under common assumptions on convexity and smoothness of the involved energies and dissipation functions, the optimization problem is equivalent to finding critical points as solutions to the corresponding variational equation, also known as D'Alembert's principle,
\begin{equation}
	\begin{split}
		&\int_{\body} \rho \ddot{\vec{u}} \cdot \delta \vec{u} \, dV+\int_{\body} \left(\frac{\partial \Psi}{\partial \tsr{C}} \colon \delta \tsr{C} 
		+\sum_{i=1}^{N} \frac{\partial \Psi}{\partial \tsr{C}_{v,i}} \colon \delta {\tsr{C}}_{v,i} 
		+ \frac{\partial \Psi}{\partial {p}}  \delta {{p}} \right)\, dV\\
		&+ \int_{\body} \sum_{i=1}^N\frac{\partial \Phi}{\partial \dot{\tsr{C}}_{v,i}} \colon \delta \tsr{C}_{v,i} \, dV
		- \int_{\body} \vec{f}\cdot \delta \vec{u} \, dV = 0 .
	\end{split}
	\label{eq:varFor}
\end{equation}
For a simplified presentation, we rewrite D'Alembert's principle in more compact form, although abusing notation concerning the variation of the dissipation function,
\begin{align}
	\int_{\body}\left(\rho \ddot{\vec{u}} \cdot \delta \vec{u}  + \delta \Psi + \delta \Phi - \vec{f}\cdot \delta \vec{u} \right)\, dV &= 0, \\
	\label{eq:dalembert_compact}
\end{align}
with,
\begin{align}
	\delta \Psi := \frac{\partial \Psi}{\partial \tsr{C}} \colon \delta \tsr{C} 
	+\sum_{i=1}^{N} \frac{\partial \Psi}{\partial \tsr{C}_{v,i}} \colon \delta {\tsr{C}}_{v,i} 
	+ \frac{\partial \Psi}{\partial {p}}  \delta {{p}},
\end{align}
and 
\begin{align}
	 \delta \Phi := \sum_{i=1}^N\frac{\partial \Phi}{\partial \dot{\tsr{C}}_{v,i}} \colon \delta \tsr{C}_{v,i}.
\end{align}
	%
	
	The variational equation \eqref{eq:varFor} implicitly defines the evolution law for $\dot{\tsr{C}}_{v,i}$ for arbitrary but fixed index $i \in \{1, \dots N\}$. We show that, for the proposed choice of dissipation function from equation \eqref{eq:def_phi}, the standard model as e.g. proposed by \cite{Haupt2013} is re-established.
	We collect all terms containing $\delta \tsr{C}_{v,i}$ from equation \eqref{eq:varFor}, and obtain, 
	\begin{equation}
		\int_\body \left( \frac{\partial \Psi_{v,i}}{\partial \tensor{C}_{v,i}} + \frac{1}{6}\eta_i \left(\tensor{C}_{v,i}^{-1} \cdot \dot{\tensor{C}}_{v,i} \cdot \tensor{C}_{v,i}^{-1}\right)\right)  \colon \delta \tensor{C}_{v,i} \, dV = 0.
		\label{eq:variDeltaCv}
	\end{equation}
	Since $\det \tsr{C}_{v,i} = 1$, we observe for its rate and variation, 
	\begin{align}
		\label{eq:CvBed1}	\frac{d}{dt} (\det \tsr{C}_{v,i}) = \tensor{C}_{v,i}^{-1} \colon \dot{\tensor{C}}_{v,i} = 0, \\
		\delta(\det\tsr{C}_{v,i}) = \tensor{C}_{v,i}^{-1} \colon \delta{\tensor{C}}_{v,i} = 0.
	\end{align}
	Thus, a contribution of the form $\gamma {\tensor{C}_{v,i}}^{-1} \colon \delta{\tensor{C}}_{v,i} = 0$ can be added without changing the equation \eqref{eq:variDeltaCv}, which leads to,
	\begin{align}
		\int_\body {\left(\frac{\partial \Psi_{v,i}}{\partial \tensor{C}_{v,i}} + \gamma \tensor{C}_{v,i}^{-1} +\frac{1}{6} \eta_i \tensor{C}_{v,i}^{-1} \cdot \dot{\tensor{C}}_{v,i} \cdot \tensor{C}_{v,i}^{-1}     \right)} \colon \delta \tensor{C}_{v,i}\, dV \nonumber = {0}.
		\label{eq: Cv_herrausgehoben}
	\end{align}
	Here, $\gamma$ is an arbitrary constant that still has to be determined in the following. 
	As $\delta \tsr C_{v,i}$ is otherwise arbitrary, the equality holds in classical sense,
	 \begin{equation}
		 	\label{eq:NLVFV1}\frac{\partial \Psi_{v,i}}{\partial \tensor{C}_{v,i}} + \gamma \tensor{C}_{v,i}^{-1} +\frac{1}{6} \eta_i \tensor{C}_{v,i}^{-1} \cdot \dot{\tensor{C}}_{v,i} \cdot \tensor{C}_{v,i}^{-1}     = \tsr{0},
		 \end{equation}
	  which directly implies, 
	 \begin{equation}
		 	\label{eq:NLVFV3}
		 	\dot{\tensor{C}}_{v,i} = -\frac{6}{\eta_i} \tensor{C}_{v,i} \cdot \left(\frac{\partial \Psi_{v,i}}{\partial \tensor{C}_{v,i}} + \gamma \tensor{C}_{v,i}^{-1}\right) \cdot \tensor{C}_{v,i}.	
		 \end{equation}
	Contraction with $\tensor{C}_{v,i}^{-1}$ leads through equation \eqref{eq:CvBed1} to,
	\begin{align}
		0 = \tensor{C}_{v,i}^{-1} : \left[ \tensor{C}_{v,i} \cdot \left(\frac{\partial \Psi_{v,i}}{\partial \tensor{C}_{v,i}} + \gamma \tensor{C}_{v,i}^{-1}\right) \cdot \tensor{C}_{v,i} \right] 
		= \tensor{I} : 	\left[ \left(\tensor{C}_{v,i} \cdot \frac{\partial \Psi_{v,i}}{\partial \tensor{C}_{v,i}} + \gamma \tensor{I}\right) \cdot \tensor{C}_{v,i} \cdot  \tensor{C}_{v,i}^{-1}\right]. \nonumber 
	\end{align}
	Since $\tensor{C}_{v,i} \cdot \tensor{C}_{v,i}^{-1} = \tensor{I}$, we can simplify even further,
	\begin{align}
		0= \tensor{C}_{v,i} \cdot \frac{\partial \Psi_{v,i}}{\partial \tensor{C}_{v,i}} : \tensor{I} + \gamma \tensor{I}:\tensor{I}   = -\tensor{M}_{v,i}:\tensor{I} + 3 \gamma. \nonumber
\end{align}
	Above, we introduce $\tsr M_{v,i}$ as the $i^\text{th}$ Mandel-type stress tensor,
	\begin{align}
		\tsr M_{v,i} = -\tsr C_{v,i} \cdot \frac{\partial \Psi_{v,i}}{\partial \tsr C_{v,i}}.
	\end{align} 
	With this, we finally have an expression to determine the unknown parameter $\gamma$,
	\begin{align}
		\gamma = \frac{1}{3}\opTrace\tensor{M}_{v,i}.
	\end{align}
We can now substitute $\gamma$ back into equation \eqref{eq:NLVFV3} and after a few algebraic manipulations we get,
\begin{align}
	\label{eq:devM1}
	\dot{\tensor{C}}_{v,i} 
	&=\frac{6}{\eta_i}\opdev(\tensor{M}_{v,i})  \cdot \tensor{C}_{v,i},
\end{align}
which is stated in similar form by \cite{ask2012electrostriction} for viscoelastic electrostrictive polymers.
To compare the final evolution law to standard theory, we explicitly compute the Mandel-type stresses for the viscoelastic potential $\Psi_{v,i}$ from equation \eqref{eq:psiVisco} and obtain,
\begin{align}
	\label{eq:Mvi}
	\tensor{M}_{v,i} 
	&= 
	- \frac{\mu_{v,i}}{2} \left( \hat{\tensor{C}} \cdot \tensor{C}_{v,i}^{-1}\right).
\end{align}
After combining the two equations equation \eqref{eq:devM1} and equation \eqref{eq:Mvi} we finally obtain the evolution law for $\tsr C_{v,i}$ as stated in \cite{Haupt2013},
\begin{eqnarray}
	\frac{2}{3}\eta\dot{\tensor{C}}_{v,i} =
	2\mu_{v,i} \left( \hat{\tensor{C}} - \frac{1}{3} \opTrace \left(\hat{\tensor{C}} \cdot \tensor{C}_{v,i}^{-1}\right) \tensor{C}_{v,i} \right).
\end{eqnarray}

\section{Finite elements for radially symmetric problems}
\label{sec:fem}

For radially symmetric problems, a finite element model respecting incompressibility of the viscous strains explicitely can be devised as follows. This setup is designed such that it allows for verification against various measurement setups, 
like the ball-drop experiment in the last section to determine rebound resilience.

\subsection{Spatial discretization for radially symmetric problems}
Let the body $\body$ as well as all external forces and kinematic constraints be in accordance with radial symmetry. Without restriction of generality, we consider $\vec e_z$ as axis of rotation. Deformation gradient $\tsr F$ and right Cauchy-Green  $\tsr C$ are defined in the standard way, assuming that the solution is independent of the angular coordinate $\varphi$. 

Assume $\mathcal T = \{T\}$ to be a triangular finite element mesh of the section of $\body$ under consideration.  Concerning the elastic problem, we propose to use the pair of displacement-pressure elements as proposed by Taylor and Hood (cf. \cite{TaylorHood:1973}), where both $\vec u$ and $p$ are represented by continuous basis functions $\vec N^{(u)}_i$ and $N^{(p)}_i$ respectively, and the approximation order for $\vec u$ is one higher than for $p$. 
Additionally, kinematic boundary conditions $\vec u = \vec 0$ on the boundary part $\Gamma_{fix}$ have to be satisfied. Summing up, finite element approximations for $\vec u$ and $p$ satisfy,
\begin{align}
		\vec u &\in \{ \vec v: \vec v|_T \in (P^k(T))^2, \vec v \text{ cont.}, \vec v|_{\Gamma_{fix}} = \vec 0\}, \\  
		\vec u &= \sum_{i=1}^{n_u} q_i^{(u)} \vec N^{(u)}_i, \\
		\label{eq:FEMu}
		p &\in \{ q: q|_T \in P^{k-1}(T), q \text{ cont.}\}, \\
		p &= \sum_{i=1}^{n_p} q_i^{(p)}N^{(p)}_i.
		\label{eq:FEMp}
\end{align}

In addition, it is necessary to provide a finite-dimensional approximation for the viscoelastic internal variables $\tsr C_{v,i}$. Here, it is important that the kinematic constraint of incompressibility is satisfied by construction. For a rotationally symmetric setting, this can be achieved by a non-linear representation of $\tsr C_{v,i}$ in terms of the independent viscoelastic strains $s_{i,rr}$, $s_{i,zz}$ and $s_{i,rz}$. The $i^{th}$ viscoelastic Cauchy-Green tensor respects the axisymmetric structure,
\begin{equation}
	\tensor{C}_{v,i} = 
	\begin{bmatrix}
		2s_{i,rr}+1  &  0 & 2s_{i,rz} \\
		0 & c_{i,\varphi\varphi} & 0 \\
		2s_{i,rz} & 0 & 2s_{i,zz} +1
	\end{bmatrix}.
\end{equation}
Component $c_{i,\varphi\varphi}$ is chosen such that the incompressibility constraint $\det \tsr C_{v,i} \equiv 1$ is satisfied a-priorily, which implies, 
\begin{eqnarray}
	c_{i,\varphi\varphi} = \left( (2s_{i,rr}+1)(1+2s_{i,zz}) - 4s_{i,rz}\right)^{-1} . 
\end{eqnarray} 
The viscoelastic strains $s_{i,rr},s_{i,zz}$ and $s_{i,rz}$ are defined element-wise polynomial, without continuity, one polynomial order lower than the displacements,	
\begin{align}
	\begin{split}
		s_{i,rr}, s_{i,zz}, s_{i,rz} &\in \{ s: s|_T \in P^{k-1}(T)\}, \\ 
		s_{i,\alpha\beta} &= \sum_{i=1}^{n_s} q_i^{(s,\alpha\beta)}N^{(s)}_i.
		\label{eq:FEMs}
	\end{split}
\end{align}

\subsection{Time integration}

Introducing the finite element discretizations equations \eqref{eq:FEMu}, \eqref{eq:FEMp} and \eqref{eq:FEMs} in D'Alembert's principle \eqref{eq:varFor} leads to a system of ordinary differential equations for the degree of freedom vector $\vec q$. As the physical problem includes accelerations $\ddot {\vec u}$ and viscoelastic strain rates $\dot{\tsr C}_{v,i}$ as well as the Lagrangian multiplier $p$, a differential-algebraic system of equations is generated after spatial discretization. It is well known that the Newmark-$\beta$ scheme is well-suited to solve differential-algebraic equations of index 2, where the choice of parameter $\beta$ further determines its numerical damping properties. 

We shortly present the Newmark-$\beta$ scheme adapted for the viscoelastic problem at hand. To this end, we assume that velocities $\vec v = \dot {\vec u}$ are approximated by an independent linear combination of displacement basis functions $\vec N^{(u)}_i$, via additional degrees of freedom $q^{(v)}_i$. For any time step $[t_n, t_{n+1} = t_n + \Delta t]$,
the quantities $\vec{u}_n$, $\vec v_n$, $p_n$ as well as $\tsr{C}_{v,i,n}$, $\dot{\tsr{C}}_{v,i,n}$ (or rather the viscoelastic strains $s_{i,\alpha\beta,n}$ and strain rates $\dot s_{i,\alpha\beta,n}$) at time $t_n$ are assumed to be available from the previous step.
According updates $\vec{u}$, $\vec v$, $p$, $s_{i,\alpha\beta}$, and $\dot s_{i,\alpha\beta}$ at time $t_{n+1} = t_n + \Delta t$ are to be computed, where we omit $n+1$ as an index in favor of brevity of presentation.
The Newmark-$\beta$ scheme interpreted in terms of the underlying variational equations leads to the coupled non-linear system, where we use $\delta \vec v$ as the (independent) variation of the velocity $\vec v$,
\begin{align}
	\int_{\body}&\rho\left( \vec{u} - \vec{u}_n - \Delta t\, \vec v_n \right) \cdot \delta\vec{u}\, dV
	+\left(\Delta t\right)^2\int_{\body}  \beta (\delta \Psi + \delta \Phi - \vec f \cdot \delta \vec u) dV \nonumber \\
	&+\left(\Delta t\right)^2\int_{\body}\left(\tfrac{1}{2} - \beta\right) ( \delta \Psi_n + \delta \Phi_n - \vec f_n \cdot \delta \vec u) \, dV  
	= 0,
	\label{eq:newmark1VF}\\
	\int_{\body}&\rho\left( \vec v - \vec v_n \right) \cdot \delta\vec v\, dV  
	+\Delta t\int_{\body}  \gamma \left( \left.\frac{\partial \Psi}{\partial \vec u}\right|_{t_{n+1}} \cdot \delta \vec v - \vec f \cdot \delta \vec v\right) dV \nonumber\\
	&+\Delta t\int_{\body}\left(1 - \gamma\right) \left( \left.\frac{\partial \Psi}{\partial \vec u}\right|_{t_n}\cdot \delta \vec v - \vec f_n \cdot \delta \vec v\right) \, dV = 0 .  
	\label{eq:newmark2VF}
\end{align}
Note that equation \eqref{eq:newmark1VF} does not only implement the Newmark integration scheme for the displacements, but also the incompressibility constraint via $\partial \Psi / \partial p \cdot \delta p = 0$ and the evolution law for the viscoelastic strains as indicated by the dissipation function. The variation of the dissipation function $\delta \Phi$ is to be understood in the same manner as in equation \eqref{eq:dalembert_compact}.



\section{Validation}
\label{sec:computational}
To validate the material characterization developed above, we present a comparison of computational results to measurements for a ball-drop test setup. Ball-drop tests are frequently used to determine the rebound resilience of materials by comparing the original drop height to the rebound height of the ball. When varying temperature, drop height as well as size and mass of the ball, changes in the rebound resilience are of interest.

For the purpose of computational realization we use the open-source software package \textit{Netgen/NGSolve}\footnote{\url{https://ngsolve.org}}. Via a python interface, the developed variational time stepping scheme \eqref{eq:newmark1VF}-\eqref{eq:newmark2VF} can be defined symbolically; all proposed finite element spaces are available at arbitrary order. Automatic differentiation takes care of computing correct tangent stiffnesses in a straightforward manner.

As sketched in Figure~\ref{fig:balldrop}, we consider a cylindric specimen made from Sylgard~184 with a radius of $r_S =\SI{30}{\milli \meter} $ and a height of $h_S=\SI{30}{\milli \meter} $. A steel ball of radius $r_B = \SI{15}{\milli \meter}$ and 
mass $m_B = \SI{109}{\gram}$ is dropped from an initial height of $h_0 =\SI{450}{\milli\meter}$ above the cylinder's surface.
All parameters are once again summarized in Table~\ref{tab:bps2}.

\begin{table}[h]
	\footnotesize
	\centering
	\caption{Parameters of the ball-drop test}
	\label{tab:bps2}
	\setlength{\tabcolsep}{15pt} 
	\begin{tabular}{ l c l }
		\hline
		specimen radius & $r_S$ & $\SI{30}{\milli\meter}$ \\
		\hline
		specimen height & $h_S$ & $\SI{30}{\milli\meter}$ \\
		\hline
		specimen density & $\rho_{Syl}$ & $\SI{965}{\kilogram \per \cubic \meter}$ \\
		\hline
		drop height & $h_0$ & $\SI{450}{\milli \meter}$ \\
		\hline
		ball radius & $r_B$ & $\SI{15}{\milli \meter}$ \\
		\hline	
		ball mass & $m_{B}$ & $\SI{109}{\gram}$ \\
		\hline
	\end{tabular}
\end{table}

A finite element model exploiting the model's radial symmetry is generated, where the finite element mesh consists of $223$ triangular elements. Third order elements are used in simulations, which results in a total of $2212$ displacement degrees of freedom. 
While the proposed material characterization is employed to describe the cylindric specimen, the ball is modelled as linear elastic body with Young's modulus $E = \SI{210}{\giga\pascal}$ and Poisson ratio $\nu = 0.3$. We choose a penalty-formulation for frictionless contact to describe the interaction of ball and PDMS. In experiments, the polymer's surface is powdered to reduce friction and adhesion effects to a minimum.

{\tiny \begin{figure}[h]
	\begin{center}{}{} 
		\resizebox{5cm}{!}{%
			\begin{tikzpicture}
				\draw [densely dash dot, line width=0.7pt] (0,-0.9) -- (0,5.5);
				\draw [densely dash dot, line width=0.7pt] (-0.2,4) -- (1.2,4);
				\draw [solid, line width=1pt] (0,0) -- (2,0);  
				\draw [solid, line width=1pt] (2,0) -- (2,2);	
				\draw [solid, line width=1pt] (2,2) -- (0,2);	

				\draw [solid, line width=1pt] (-0.2,-0.55) -- (0.2,-0.55);
				\draw [solid, line width=1pt] (-0.2,-0.65) -- (0.2,-0.65);
				\draw [solid, line width=1pt] (-0.2,5.25) -- (0.2,5.25);
				\draw [solid, line width=1pt] (-0.2,5.35) -- (0.2,5.35);

				\draw [solid, line width=0.5pt] (0,4) --  (1.5,5);
				\draw [-latex, line width=0.5pt] (1.5,5) --  node[midway, above right, sloped] {$r_B$}  (0.832,4+0.5546);

				\draw [-latex, line width=0.5pt] (-0.3,-1.1) -- node[above ]{$r_S$}(2,-1.1);
				\draw [solid, line width=0.5pt] (2,-0.9) -- (2,-1.3);
				\draw [latex-latex, line width=0.5pt] (2.7,0) -- node[right]{$h_S$}(2.7,2);
				\draw [solid, line width=0.5pt] (2.5,2) -- (2.9,2);
				\draw [solid, line width=0.5pt] (2.5,0) -- (2.9,0);
				\draw [latex-latex, line width=0.5pt] (2.7,2) -- node[right]{$h_0$}(2.7,3);
				
				\draw [solid, line width=0.5pt] (2.5,3) -- (2.9,3);
				
				\node at (0.4,3.5) {$m_{B}$};
				\node at (1,1) {$\rho_{Syl}$};
				
				
				\draw [-latex, line width=1.2pt] (0,0) -- node[left=2mm] {$z$}(0,0.8);
				\draw [-latex, line width=1.2pt] (0,0) -- node[below=3mm] {$x$} (0.8,0);

				\draw [solid, line width=0.5pt] (0,-0.1) -- (2,-0.1);
				\foreach \i in {0,1,...,6}{
					\draw [solid, line width = 0.5pt] (1.7-\i*0.3,-0.4) -- (2.0-\i*0.3,-0.1);	
				}
				
				\draw [solid, line width=0.5pt] (-0.7,2) -- (0,2);
				\draw [solid, line width=0.5pt] (-0.7,2.1) -- (0,2.1);
				\draw [latex-, line width=0.5pt] (-0.5,2.1) -- node[above left]{$h_{1}$}(-0.5,2.5);
				\draw [latex-, line width=0.5pt] (-0.5,2) -- (-0.5,1.5);
				\draw [solid, line width=0.5pt] (-0.5,1.5) -- (-0.5,2.5);

				\clip (0,2) rectangle (3,6);
				\draw [solid, line width=1pt] (0,4) circle(1);
				
				\clip (0,2) rectangle (4,2.5);
				\draw [dashed, line width=1pt] (0,3.1) circle(1);

			\end{tikzpicture}
			
		}
	\end{center}
	\caption{Computational setup of the ball-drop test.}
	\label{fig:balldrop}
	
\end{figure}
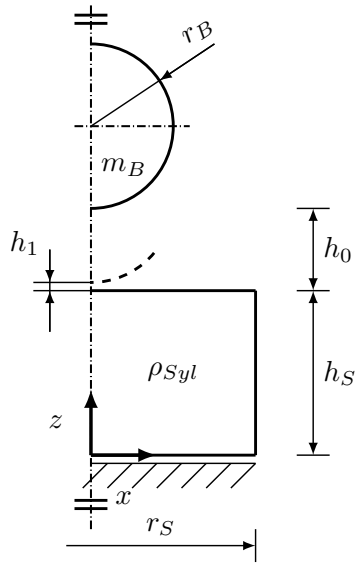}

\begin{figure}[h]
	\begin{center}{}{} 
		\resizebox{7cm}{!}{%
			\begin{tikzpicture}
				\draw[-latex] (0,0) -- (5,0) node[below] {$t$};
				\draw[-latex] (0,0) -- (0,3) node[left] {$h$};
				
				\draw[black, thick] (0,2) parabola (2,0);
				
				\draw[black, thick] (3,1.2) parabola (2,0); 
				\draw[black, thick] (3,1.2) parabola (4,0);
				\draw[black] (0,0) -- (0,-0.3) node[below] {$t_0$};
				\draw[black] (1.9,0.2) -- (1.9,-0.3) node[below left] {$t_1$};
				\draw[black] (2.1,0.2) -- (2.1,-0.3) node[below right] {$t_2$};
				\draw[black] (0,2) -- (-0.3,2) node[left] {$h_0$};
				\draw[black] (1.9,0.2) -- (1.7,0.2) node[left] {$h_1, v_1$};
				\draw[black] (2.1,0.2) -- (2.3,0.2) node[right] {$h_2, v_2$};
				\draw[black] (3,1.2) -- (3.5,1.2) node[right] {$h_{r,1}$};
			\end{tikzpicture}
		}
	\end{center}
	\caption{Idealized trajectory of the ball.}
	\label{fig:traj}
\end{figure}
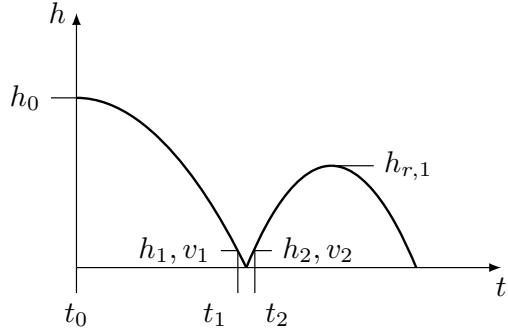

Assume $t_0 = \SI{0}{\second}$ to denote the time at which the ball is released. 
The actual simulation is restricted to the time interval around the ball's first impact. To be concise, time stepping is initialized for the ball at height $h_{1} = \SI{0.02}{\meter}$, which is close to the specimen's surface. This results in an initial time $t_1$ an initial velocity $\vec v_1 = v_1 \vec e_z$ set to,
\begin{align}
	t_1 &= \sqrt{\frac{2}{g} (h_0 - h_{1})}, & v_1 &= - g  t_1.
\end{align}
In Figure~\ref{fig:traj}, these time instants as well as the corresponding velocities and observed heights are visualized.

In simulations, the Newmark parameters are chosen as $\beta = 0.25$ and $\gamma=0.5$ to avoid numerical damping. The time step size is set to to $\Delta t = \SI{1e-4}{\second}$ to generate accurate results when using the second-order accurate Newmark-$\beta$ method. 
The simulation is stopped once the ball exceeds the initial distance of $h_{1}$, let this time instant be denoted as $t_2$. The first rebound height $h_{r,1}$ can then be computed from height $h_2$ and vertical velocity $v_2$ at time $t_2$,
\begin{equation}
	h_{r,1} = h_2 + \frac{g}{2} v_2^2 .
\end{equation}

\begin{figure}[H]
	\centering
	\includegraphics[width=0.69\linewidth]{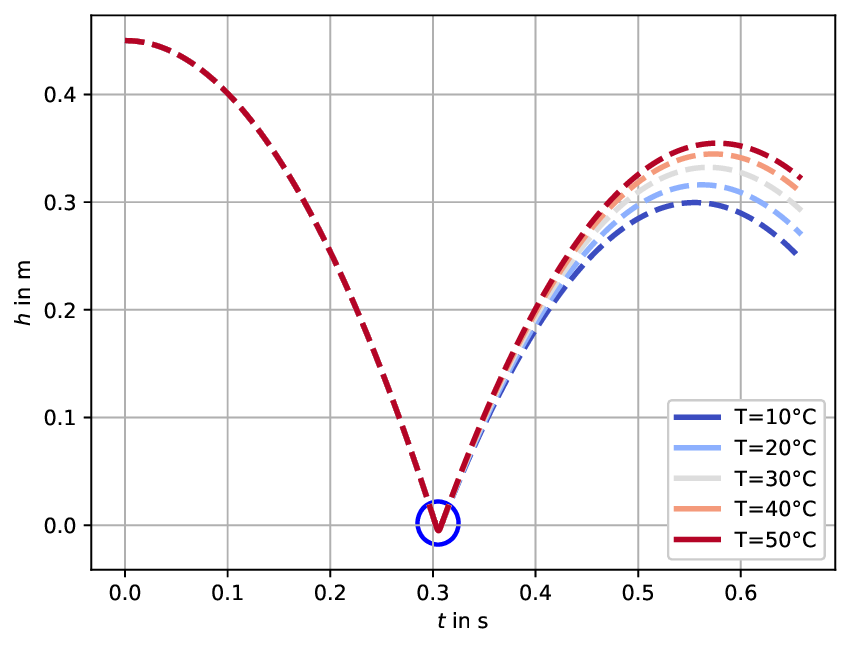}
	\caption{Ball-drop trajectory, $h_0 = \SI{0.45}{\meter}, r_B = \SI{15}{\milli\meter} $, using material data for $N=3$ from Table~\ref{tab:paraO3}.}
	\label{fig:bdh0}
\end{figure}

\begin{figure}[H]
	\centering
	\includegraphics[width=0.69\linewidth]{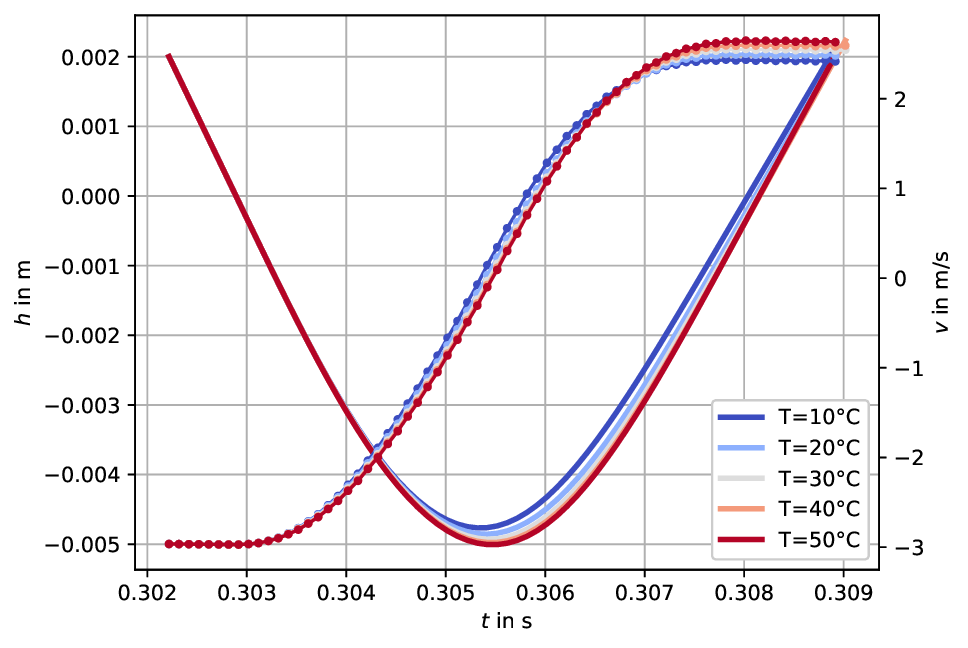}
	\caption{Impact trajectory (cmp.~Figure~\ref{fig:bdh0}), solid lines indicate the ball's trajectory, dots mark its velocity.}
	\label{fig:zoombdh0}
\end{figure}

In Figure~\ref{fig:bdh0}, the ball's computed trajectory at different temperatures is presented, where the third order material characterization from Table~\ref{tab:paraO3} has been used. Five different temperatures were chosen such that they cover the range of DTMA measurements, we use $10, 20, 30, 40 \text{ and } \SI{50}{\degreeCelsius}$ respectively. 
Free fall and rebound trajectories above height $h_{1}$ and $h_{2}$ respectively were generated analytically and are identified as dashed lines. Results of the actual simulation are highlighted in blue, and presented in detail in Figure~\ref{fig:zoombdh0}. In addition to the trajectories, the ball's vertical velocity is displayed for all temperatures. An increase in the rebound height for increasing temperatures can be clearly observed. 

Damping properties of the material are often characterized through the rebound resilience or loss factor in the ball drop test, see \cite{Emminger:2021}. The loss factor is given by $\tan \delta = h_{r,1} / h_0$, often also referred to as $R$.
Both in the experiment and in the simulation, a number of parameters can be changed in order to better study the behaviour.
For example, the variation of the original height $h_0$, the ball's radius $r_B$, the actual temperature as well as the order of the material model can provide suitable information. 

\begin{figure}[H]
	\centering
	\includegraphics[width=0.69\linewidth]{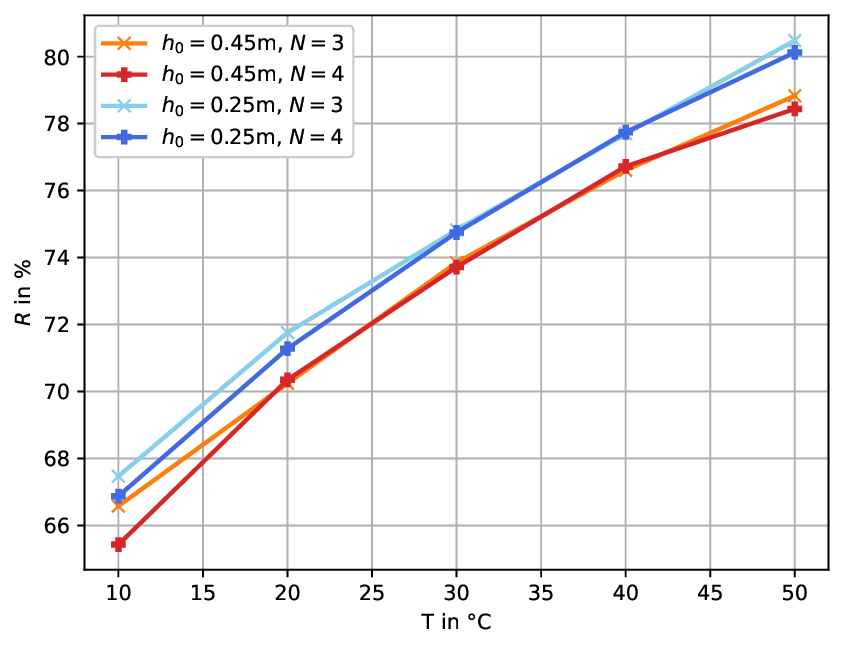}
	\caption{Rebound resilience comparison.}
	\label{fig:rebresvglo3o4}
\end{figure}

Figure~\ref{fig:rebresvglo3o4} shows the variation of the rebound resilience for different temperatures. From these results, we see that the third and fourth order material model lead to similar results; thus, a restriction to third order is viable to save computational effort.
On the other hand, we observe that by decreasing the original drop height to $\SI{0.25}{\meter}$, an increase in the rebound resilience is achieved. 
For increasing temperatures, the resilience increases independently of the drop height. At higher temperature, the material acts less viscous. 

\begin{figure}[H]
	\centering
	\includegraphics[width=0.69\linewidth]{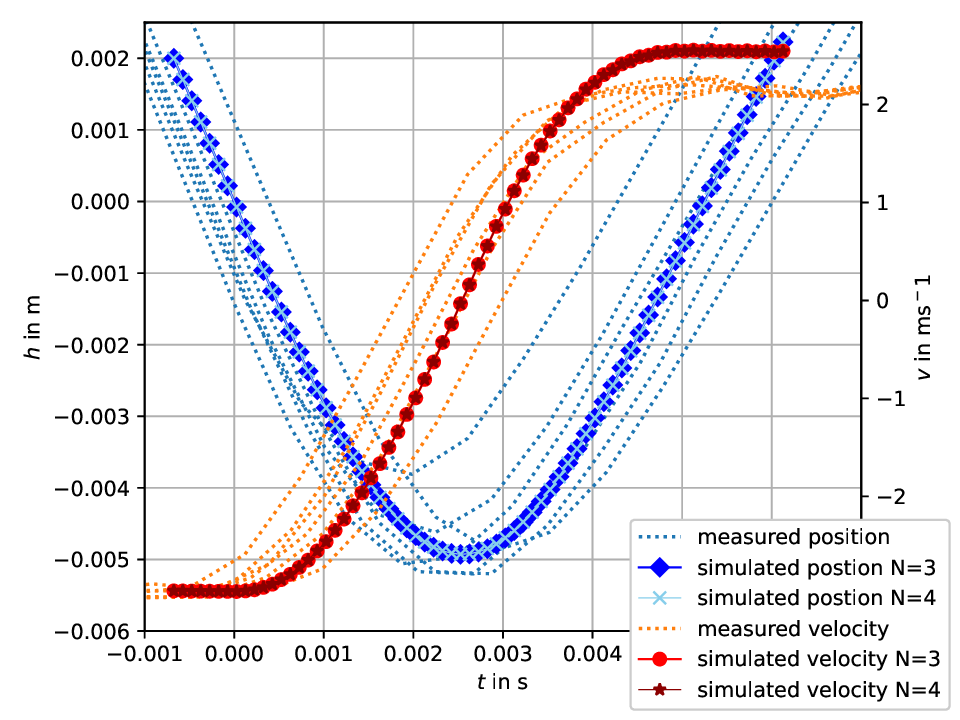}
	\caption{Ball-drop measurements for $ h_0 = \SI{0.45}{\meter}, r_B = \SI{15}{\milli \meter}$ at $T=\SI{30}{\degreeCelsius}$.}	\label{fig:balldroph45t30d30}
\end{figure}

The computational results are compared to measurements taken in an according setup.
For this purpose, an experimental setup has been designed, where the ball is released from an electromagnetic mount. A high-speed optical camera was used to track the trajectory of the ball.
Care was taken to ensure that the surface of the sample was neatly powdered before each measurement in order to exclude possible friction and adhesion effects. 
Several of these measurements were carried out and are shown as dotted lines in Figure~\ref{fig:balldroph45t30d30}. 
The trajectories obtained from simulations, as shown in Figure~\ref{fig:zoombdh0}, are compared to these measurement data curves around the time of impact.
Again, third and fourth order show no difference for the height course and also the velocity.

It shows that the depth of penetration observed in simulation and experiment fit closely. In simulations, however, a higher rebound resilience of $R = \SI{72.6}{\percent}$ is observed as compared to $R = \SI{52.9}{\percent}$ in the experiments. Consistently, the velocity of the ball after impact is higher. It is not clear where the discrepancy originates from; certainly the friction-less contact model, the assumption of perfect rotational symmetry in displacements and also the rigid bedding of the PDMS specimen add to the computed rebound height, as dissipation is inhibited.

\begin{figure}[h]
	\centering
	\includegraphics[width=0.69\linewidth]{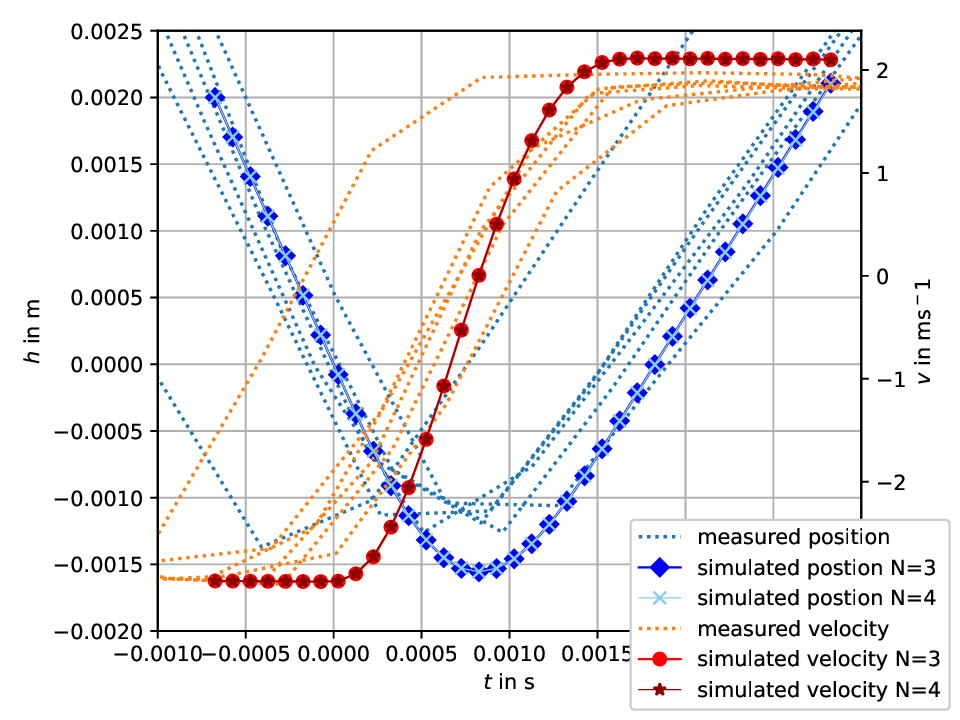}
	\caption{Ball-drop measurements for $ h_0 = \SI{0.45}{\meter}, r_B = \SI{5}{\milli \meter}$ at $T=\SI{30}{\degreeCelsius}$.}	\label{fig:balldroph45t30d10}
\end{figure}

In a second set of experiments, a ball of radius $r_B = \SI{5}{\mm}$ was dropped again from the original drop height of $h_0 = \SI{0.45}{\meter}$, the ambient temperature is at $T = \SI{30}{\degreeCelsius}$. 
In Figure~\ref{fig:balldroph45t30d10}, corresponding measurements and simulation results are compared. Again, correspondence is observed for the indentation depth. Concerning rebound resilience and velocity after impact, the results fit better for this smaller ball. The rebound resilience in simulation is found at $R=\SI{50.38}{\percent}$, whereas the measured rebound resilience was determined as $R = \SI{42.8}{\percent}$


\section{Conclusion}
In the present article, we have introduced a thermodynamically consistent characterization for viscoelastic materials for modeling the behavior of the PDMS Sylgard~184. To determine the model's parameters, a dynamical temperature mechanical analysis was conducted in a limited frequency range. The time-temperature superposition principle was used successfully to generate an extended frequency response.
For this data, Prony-parameters based on the Wiechert model were identified using a non-negative least squares fitting algorithm. 
To verify this constitutive model, a mixed finite element model for rotationally symmetric problems was created.
A ball-drop test setup was designed in order to conduct measurements on the rebound resilience for different ball sizes and drop heights. Data generated in the finite element simulation were then compared to optical tracking data from the experiment.
Our numerical results match the measured data almost perfectly as far as the indentation depth is concerned. Rebound resiliences differ, however, as we observe less dissipation in computations than in experiments.
To analyse these deviations in detail, the effects of friction-less contact and the imposed radial symmetry in computations need to be quantified in the future. Additionally, DTMA measurements in an even larger range of temperature may lead to a better characterization of the viscous behaviour at high frequencies, which are certainly excited in a ball-drop test.

%
%
%
%
%
%

\bibliographystyle{plain}
\bibliography{References}

\begin{thebibliography}{10}

\bibitem{ariati2021polydimethylsiloxane}
Ronaldo Ariati, Flaminio Sales, Andrews Souza, Rui~A Lima, and Jo{\~a}o
  Ribeiro.
\newblock Polydimethylsiloxane composites characterization and its
  applications: a review.
\newblock {\em Polymers}, 13(23):4258, 2021.

\bibitem{ask2012electrostriction}
Anna Ask, Andreas Menzel, and Matti Ristinmaa.
\newblock Electrostriction in electro-viscoelastic polymers.
\newblock {\em Mechanics of Materials}, 50:9--21, 2012.

\bibitem{borok2021pdms}
Alexandra Bor{\'o}k, Krist{\'o}f Laboda, and Attila Bony{\'a}r.
\newblock Pdms bonding technologies for microfluidic applications: A review.
\newblock {\em Biosensors}, 11(8):292, 2021.

\bibitem{Brinson2015}
Hal~F. Brinson and L.~Catherine Brinson.
\newblock {\em Polymer Engineering Science and Viscoelasticity - An
  Introduction}.
\newblock Springer, Berlin, Heidelberg, 2015.

\bibitem{Cardoso:2018}
C\'atia Cardoso, Carla~S. Fernandes, Rui Lima, and {Jo\~{a}o} Ribeiro.
\newblock Biomechanical analysis of pdms channels using different hyperelastic
  numerical constitutive models.
\newblock {\em Mechanics Research Communications}, 90:26--33, 2018.

\bibitem{CLARSON1985930}
S.J Clarson, K~Dodgson, and J.A Semlyen.
\newblock Studies of cyclic and linear poly(dimethylsiloxanes): 19. glass
  transition temperatures and crystallization behaviour.
\newblock {\em Polymer}, 26(6):930--934, 1985.

\bibitem{Deguchi:2015}
Shinji Deguchi, Junya Hotta, Sho Yokoyama, and Tsubasa~S Matsui.
\newblock Viscoelastic and optical properties of four different pdms polymers.
\newblock {\em Journal of Micromechanics and Microengineering}, 25(9):097002,
  2015.

\bibitem{dewimille2005synthesis}
Ludivine Dewimille, Bruno Bresson, and Liliane Bokobza.
\newblock Synthesis, structure and morphology of poly (dimethylsiloxane)
  networks filled with in situ generated silica particles.
\newblock {\em Polymer}, 46(12):4135--4143, 2005.

\bibitem{Doghri2013}
Issam Doghri.
\newblock {\em Mechanics of Deformable Solids - Linear, Nonlinear, Analytical
  and Computational Aspects}.
\newblock Springer Science and Business Media, Berlin Heidelberg, 2013.

\bibitem{eduok2017recent}
Ubong Eduok, Omar Faye, and Jerzy Szpunar.
\newblock Recent developments and applications of protective silicone coatings:
  A review of pdms functional materials.
\newblock {\em Progress in Organic Coatings}, 111:124--163, 2017.

\bibitem{Emminger:2021}
Carina Emminger, Umut~D {\c{C}}akmak, Rene Preuer, Ingrid Graz, and Zolt{\'a}n
  Major.
\newblock Hyperelastic material parameter determination and numerical study of
  {T}{P}{U} and {P}{D}{M}{S} dampers.
\newblock {\em Materials}, 14(24):7639, 2021.

\bibitem{Ferry1980}
John~D. Ferry.
\newblock {\em Viscoelastic Properties of Polymers -}.
\newblock John Wiley \& Sons, New York, 1980.

\bibitem{Haupt2013}
Peter Haupt.
\newblock {\em Continuum Mechanics and Theory of Materials}.
\newblock Springer Science and Business Media, Berlin Heidelberg, 2013.

\bibitem{Klompen}
ETJ Klompen and LE~Govaert.
\newblock Nonlinear viscoelastic behaviour of thermorheologically complex
  materials: A modelling approach.
\newblock {\em Mechanics of Time-dependent Materials - MECH TIME-DEPEND MATER},
  3, 01 1999.

\bibitem{knapp2021controlling}
Andr{\'e} Knapp, Lisa~Julia Nebel, Mirko Nitschke, Oliver Sander, and Andreas
  Fery.
\newblock Controlling line defects in wrinkling: a pathway towards hierarchical
  wrinkling structures.
\newblock {\em Soft Matter}, 17(21):5384--5392, 2021.

\bibitem{kraus2017parameter}
Michael~A Kraus, Miriam Schuster, Johannes Kuntsche, Geralt Siebert, and Jens
  Schneider.
\newblock Parameter identification methods for visco-and hyperelastic material
  models.
\newblock {\em Glass Structures \& Engineering}, 2(2):147--167, 2017.

\bibitem{KunzemannEtal:2023}
Mario Kunzemann, Astrid Pechstein, and Alexander Humer.
\newblock Thermodynamically consistent modelling of dielectric viscoelastic
  solids under finite strain.
\newblock In D.A. Saravanos, A.~Benjeddou, N.~Chrysochoidis, and T.~Theodosiou,
  editors, {\em Proceedings of the X ECCOMAS Thematic Conference on Smart
  Structures and Materials}, 2023.

\bibitem{LeeEtal:2021}
Hoo~Min Lee, Jaebum Sung, Byeongjo Ko, Heewon Lee, Sangyeun Park, Hongyun So,
  and Gil~Ho Yoon.
\newblock Modeling and application of anisotropic hyperelasticity of pdms
  polymers with surface patterns obtained by additive manufacturing technology.
\newblock {\em Journal of the Mechanical Behavior of Biomedical Materials},
  118:104412, 2021.

\bibitem{LI2006580}
Wanwan Li, Feng Liu, Liuhe Wei, and Tong Zhao.
\newblock Synthesis, morphology and properties of polydimethylsiloxane-modified
  allylated novolac/4,4-bismaleimidodiphenylmethane.
\newblock {\em European Polymer Journal}, 42(3):580--592, 2006.

\bibitem{LinEtal:2009}
I-Kuan Lin, Kuang-Shun Ou, Yen-Ming Liao, Yan Liu, Kuo-Shen Chen, and Xin
  Zhang.
\newblock Viscoelastic characterization and modeling of polymer transducers for
  biological applications.
\newblock {\em Journal of Microelectromechanical systems}, 18(5):1087--1099,
  2009.

\bibitem{LOMELLINI19924983}
Paolo Lomellini.
\newblock Williams-landel-ferry versus arrhenius behaviour: polystyrene melt
  viscoelasticity revised.
\newblock {\em Polymer}, 33(23):4983--4989, 1992.

\bibitem{Long:2017}
Kevin~Nicholas Long and Judith~Alice Brown.
\newblock A linear viscoelastic model calibration of sylgard 184.
\newblock Technical report, Sandia National Lab.(SNL-NM), Albuquerque, NM
  (United States), 2017.

\bibitem{miehe2011variational}
C~Miehe, D~Rosato, and B~Kiefer.
\newblock Variational principles in dissipative electro-magneto-mechanics: a
  framework for the macro-modeling of functional materials.
\newblock {\em International Journal for Numerical Methods in Engineering},
  86(10):1225--1276, 2011.

\bibitem{Miehe:2002}
Christian Miehe.
\newblock Strain-driven homogenization of inelastic microstructures and
  composites based on an incremental variational formulation.
\newblock {\em International Journal for numerical methods in engineering},
  55(11):1285--1322, 2002.

\bibitem{MieheEtal:2002}
Christian Miehe, Nikolas Apel, and Matthias Lambrecht.
\newblock Anisotropic additive plasticity in the logarithmic strain space:
  modular kinematic formulation and implementation based on incremental
  minimization principles for standard materials.
\newblock {\em Computer methods in applied mechanics and engineering},
  191(47-48):5383--5425, 2002.

\bibitem{Nunes:2011}
L.C.S. Nunes.
\newblock Mechanical characterization of hyperelastic polydimethylsiloxane by
  simple shear test.
\newblock {\em Materials Science and Engineering: A}, 528(3):1799--1804, 2011.

\bibitem{TaylorHood:1973}
Cedric Taylor and Paul Hood.
\newblock A numerical solution of the {N}avier-{S}tokes equations using the
  finite element technique.
\newblock {\em Computers \& Fluids}, 1(1):73--100, 1973.

\bibitem{Tschoegl2012}
Nicholas~W. Tschoegl.
\newblock {\em The Phenomenological Theory of Linear Viscoelastic Behavior - An
  Introduction}.
\newblock Springer Science \& Business Media, Berlin Heidelberg, 2012.

\bibitem{Tschoegl2002TheEO}
Nicholas~W. Tschoegl, Wolfgang Knauss, and Igor Emri.
\newblock The effect of temperature and pressure on the mechanical properties
  of thermo- and/or piezorheologically simple polymeric materials in
  thermodynamic equilibrium – a critical review.
\newblock {\em Mechanics of Time-Dependent Materials}, 6:53--99, 2002.

\bibitem{zaman2019comprehensive}
Qaiser Zaman, Khalid~Mahmood Zia, Mohammad Zuber, Yahia~Nasser Mabkhot, Faisal
  Almalki, and Taibi~Ben Hadda.
\newblock A comprehensive review on synthesis, characterization, and
  applications of polydimethylsiloxane and copolymers.
\newblock {\em International Journal of Plastics Technology}, 23:261--282,
  2019.

\bibitem{zhu2017recent}
Deyong Zhu, Stephan Handschuh-Wang, and Xuechang Zhou.
\newblock Recent progress in fabrication and application of
  polydimethylsiloxane sponges.
\newblock {\em Journal of Materials Chemistry A}, 5(32):16467--16497, 2017.

\end{thebibliography}

\end{document}